\documentclass[11pt,a4paper]{article}

\usepackage[utf8]{inputenc}
\usepackage[T1]{fontenc}
\usepackage[english]{babel}
\usepackage{newpxtext}
\usepackage{amsmath}
\usepackage{newpxmath}
\usepackage{microtype}
\usepackage{tikz}
\usetikzlibrary{arrows.meta,positioning}

\usepackage[
  top=2.6cm,
  bottom=2.6cm,
  left=2.7cm,
  right=2.7cm,
  headheight=15pt,
  headsep=0.7cm,
  footskip=1.2cm
]{geometry}

\usepackage{setspace}
\usepackage[
  indent=0pt,
  skip=0.65\baselineskip plus 2pt minus 1pt
]{parskip}

\usepackage{xcolor}
\definecolor{accent}{HTML}{163A5F}
\definecolor{linkblue}{HTML}{245A81}
\definecolor{muted}{HTML}{59636E}
\definecolor{rulegray}{HTML}{D4DBE2}

\usepackage{titlesec}

\titleformat{\section}
  {\normalfont\large\bfseries\color{accent}}
  {\thesection}{0.7em}{}

\titleformat{\subsection}
  {\normalfont\normalsize\bfseries}
  {\thesubsection}{0.7em}{}

\titleformat{\subsubsection}
  {\normalfont\normalsize\itshape}
  {\thesubsubsection}{0.7em}{}

\titlespacing*{\section}
  {0pt}{1.6\baselineskip}{0.55\baselineskip}

\titlespacing*{\subsection}
  {0pt}{1.1\baselineskip}{0.35\baselineskip}

\titlespacing*{\subsubsection}
  {0pt}{0.85\baselineskip}{0.25\baselineskip}

\usepackage{graphicx}
\usepackage{booktabs}
\usepackage{tabularx}
\usepackage{array}
\usepackage{enumitem}

\setlist{
  topsep=0.35\baselineskip,
  itemsep=0.2\baselineskip,
  parsep=0pt,
  leftmargin=*
}

\usepackage{caption}
\renewcommand{\arraystretch}{1.15}

\numberwithin{equation}{section}

\usepackage[authoryear,round]{natbib}
\usepackage{xurl}
\usepackage[
  colorlinks=true,
  linkcolor=linkblue,
  citecolor=linkblue,
  urlcolor=linkblue,
  bookmarksnumbered=true,
  pdfdisplaydoctitle=true,
  pdftitle={Bayesian Selection of Edge Spectral Modes for Crash Counts
    on Urban Road Networks with Applications in Barcelona and Bogotá},
  pdfauthor={Danna L. Cruz-Reyes, Juan Sosa, Carlos A. Martínez},
  pdfsubject={Bayesian spatial modeling of road-network crash counts},
  pdfkeywords={Bayesian variable selection, crash frequency,
    road network, spectral filtering, spike-and-slab prior}
]{hyperref}

\usepackage{bookmark}
\usepackage[nameinlink,noabbrev]{cleveref}
\usepackage{fancyhdr}
\fancypagestyle{firstpage}{
  \fancyhf{}
  \fancyfoot[C]{\small\thepage}

}

\renewenvironment{abstract}
  {%
    \par\addvspace{0.8\baselineskip}
    \begingroup
    \small
    \setstretch{1.08}
    \noindent{\normalsize\bfseries\color{accent}Abstract}\par
    \vspace{0.2\baselineskip}
    \ignorespaces
  }
  {%
    \par
    \endgroup
    \addvspace{0.5\baselineskip}
  }

\begin{document}

\thispagestyle{firstpage}

% ------------------------------------------------------------
% Title and authors
% ------------------------------------------------------------
\begin{center}

{\fontsize{19}{23}\selectfont\bfseries\color{accent}
Bayesian Selection of Edge Spectral Modes\\
for Crash Counts on Urban Road Networks\\
with Applications in Barcelona and Bogotá\par}

\vspace{0.8cm}

\begin{center}
\begin{tabular}{@{}c@{\hspace{1em}}c@{\hspace{1em}}c@{}}
{\Large Danna L. Cruz-Reyes\textsuperscript{1,*}} &
{\Large Juan Sosa\textsuperscript{1}} &
{\Large Carlos A. Martínez\textsuperscript{2}} \\[0.3em]
{\small\href{mailto:dlcruzr@unal.edu.co}{dlcruzr@unal.edu.co}} &
{\small\href{mailto:jcsosam@unal.edu.co}{jcsosam@unal.edu.co}} &
{\small\href{mailto:camartinezn@unal.edu.co}{camartinezn@unal.edu.co}}
\end{tabular}
\end{center}

\vspace{0.4cm}

{\small\color{muted}
\textsuperscript{1}Departamento de Estadística,
Universidad Nacional de Colombia, Bogotá, Colombia\par
\textsuperscript{2}Departamento de Producción Animal,
Universidad Nacional de Colombia, Bogotá, Colombia\par}

{\footnotesize\color{muted}
\textsuperscript{*}Corresponding author.
\par}

% Optional version date:
% \vspace{0.3cm}
% {\small\color{muted}\today\par}

\end{center}

\vspace{0.3cm}
{\color{rulegray}\hrule height 0.5pt}

% ------------------------------------------------------------
% Abstract and keywords
% ------------------------------------------------------------
\begin{abstract}
Crash counts on urban road networks exhibit spatial dependence shaped by road connectivity rather than Euclidean proximity alone. We propose a Bayesian negative-binomial model that represents spatial variation through a parsimonious combination of spectral components defined directly on road segments. The model combines the normalized random edge neighborhood Gaussian (RENeGe) operator with a continuous spike-and-slab prior whose mixture indicators are marginalized out. Eigenvalue-dependent slab variances link spectral regularization to network-supported smoothness, while posterior slab-membership probabilities quantify support for individual components and uncertainty in effective spectral complexity. Segment length provides a provisional exposure offset, so fitted rates describe crash frequency per unit road length rather than traffic-adjusted risk. A simulation study comprising 1,200 datasets and 4,800 Bayesian fits yielded the lowest mean spatial-field recovery error and highest mean test log predictive density for the proposed model under sparse RENeGe truth. Under the alternative generating mechanisms considered, its predictive performance remained close to that of dense spectral models and consistently exceeded that of a nonspatial model. Applications to Barcelona and central Bogot\'a identified distinct posterior spectral patterns. Exploratory leave-one-segment-out comparisons showed practically indistinguishable predictive accuracy for the proposed model and a low-rank edge conditional autoregressive (CAR) model, both of which outperformed a nonspatial negative-binomial model; A spectral Besag–York–Mollié 2 (BYM2) model \cite{riebler2016intuitive}, combining a network-structured spatial component with segment-specific unstructured heterogeneity, achieved the highest exploratory predictive score. These findings support probabilistic summaries of spectral complexity without a detectable predictive penalty relative to the corresponding CAR model within the same candidate basis. However, simulation convergence limitations and a data-dependent intercept prior constrain performance claims, while substantive cross-city comparisons require verified data provenance, harmonized observation periods, and spatially blocked predictive validation.
\end{abstract}

{\small
\noindent\textbf{Keywords:}
Bayesian variable selection; crash frequency; edge-based spatial modeling;
road networks; spectral filtering; spike-and-slab prior.
\par}

\vspace{0.4cm}
{\color{rulegray}\hrule height 0.5pt}

% ============================================================
% MAIN TEXT
% ============================================================

\section{Introduction}
\label{sec:introduction}

Road crashes occur on transportation networks, where spatial dependence is shaped by road connectivity as well as geographic proximity. Intersections and road segments may share traffic conditions, geometric characteristics, land-use environments, and unmeasured determinants of crash frequency. Neglecting this dependence can compromise uncertainty quantification and estimation efficiency; bias may also arise when omitted spatial factors are associated with included covariates. Bayesian spatial models, particularly conditional autoregressive (CAR) models, provide a widely used framework for addressing these features in road-safety analysis \citep{song2006,aguerovalverde2008,guo2010,xie2014,xu2017}.

An established strand of this literature defines the spatial process on intersections. Bayesian CAR models have been used to represent dependence among signalized intersections along common corridors \citep{guo2010} and among closely spaced intersections in dense urban networks \citep{xie2014}. Other formulations combine spatially structured and unstructured intersection-level random effects to accommodate spatial dependence and residual heterogeneity \citep{mitra2009}. In these models, the latent field is indexed by intersections, with information shared according to a nodal adjacency or distance structure. This support is appropriate when the response consists of intersection-level crash counts. When crashes are instead recorded on road segments, an intersection-level latent field requires a mapping to the observational support. For a segment $e=(i,j)$, possible mappings include the endpoint average $s_e=(u_i+u_j)/2$, a sum, or an exposure-weighted combination, where $u_i$ and $u_j$ denote the endpoint effects. The induced segment field consequently depends on both the nodal model and the chosen mapping.

Road segments are themselves natural observational units for crash-frequency analysis. Spatial effects have been defined directly on neighboring segments \citep{aguerovalverde2008}, and joint Bayesian models have been developed for segment and intersection crash counts \citep{zenghuang2014}. Joint formulations can represent dependence between these two types of units in addition to dependence within each type. These approaches motivate aligning the latent spatial field with the support of the observed response: when counts are recorded on segments, a segment-level field represents their dependence directly, without requiring a projection from intersection-level effects.

The random edge neighborhood Gaussian (RENeGe) model provides a framework for this construction \citep{cruzreyes2023spasta}. It defines dependence among road segments through shared endpoints. Equivalently, the road network induces a line graph in which each vertex represents a road segment and two vertices are adjacent when their corresponding segments share an intersection. The eigensystem of the normalized edge operator supplies network-supported patterns of spatial variation. Their spectral interpretation depends on the operator and its normalization; they describe variation with respect to network connectivity rather than Euclidean distance alone.

We use these edge spectral components to construct a Bayesian model for overdispersed crash counts. A continuous spike-and-slab prior distinguishes strongly shrunk coefficients from those assigned to a more diffuse component, with eigenvalue-dependent slab variances linking regularization to network-supported smoothness. Because both mixture components are continuous, the prior does not set coefficients exactly to zero. Instead, posterior slab-membership probabilities summarize support for candidate components and quantify uncertainty in effective spectral complexity. The resulting representation combines estimation of the segment-level spatial field with probabilistic assessment of its spectral contributions.

We compare the proposed model with two alternatives defined on the same road-segment support and using the same retained RENeGe basis. A spectral CAR model retains all candidate components without a selection mixture, whereas a spectral BYM2 model combines the structured field with segment-specific unstructured heterogeneity. Holding the support and candidate basis fixed provides a controlled comparison of mixture-based regularization, dense spectral regularization, and additional local heterogeneity. Here, ``dense'' refers to retaining all components of the truncated candidate basis, rather than fitting a full-rank spatial model. A nonspatial negative-binomial model provides an additional baseline.

The empirical analysis considers Barcelona and central Bogot\'a. In both applications, segment-level crash counts follow a negative-binomial likelihood, with the logarithm of segment length included as a provisional exposure offset. Consequently, fitted rates describe crash frequency per unit road length rather than traffic-adjusted crash risk. Within each city, we compare models on a common segment support, examine posterior spatial fields and their leading spectral contributions, and assess predictive performance. A complementary simulation study uses both empirical network geometries to generate sparse RENeGe, dense CAR, BYM2, and localized-hotspot fields. These scenarios assess field recovery, component identification, prediction, and computational performance under different forms of spatial structure.

The methodological contribution is the integration of segment-supported edge geometry, eigenvalue-informed continuous-mixture regularization, and posterior summaries of spectral complexity for overdispersed counts. The common-basis comparisons assess the consequences of this regularization relative to retaining all candidate components and adding unstructured heterogeneity. Spatial-field decompositions and posterior slab-membership probabilities further support interpretation of the fitted network patterns, while the simulation and applications delineate the conditions under which this parsimonious representation is useful.

\subsection{Position within spectral spatial modeling}

Eigenvector spatial filtering represents spatial dependence through graph-derived basis functions, with established fixed- and random-effect formulations \citep{griffith2003,griffith2019,murakami2015}. Graph signal processing similarly interprets graph spectra as network-adapted representations of variation \citep{shuman2013}. These perspectives motivate the use of edge spectral components, although transformations applied during basis construction require care: centering an eigenvector generally does not preserve its eigenvector property. Accordingly, the centered columns used here are understood as spectral basis functions derived from the edge operator, and their associated eigenvalues refer to the original eigensystem.

Bayesian spatial dimension reduction and projection methods also address the interaction between covariate effects and spatial structure \citep{hughes2013}. Hierarchical shrinkage has been applied to Moran eigenvector filters; for example, \citet{donegan2020} use a regularized horseshoe prior to quantify coefficient uncertainty without conditioning on a single stepwise-selected model. Bayesian spectral regularization and eigenvector selection therefore have established precedents. The present contribution combines these broader ideas with edge-neighborhood geometry and an eigenvalue-informed continuous-mixture prior for segment-level crash counts, complementing existing segment and joint segment--intersection models \citep{aguerovalverde2008,zenghuang2014,xu2017}. The comparisons assess interpretable probabilistic selection within a fixed candidate basis; they do not establish superiority over full-rank spatial models, alternative shrinkage priors, or the broader range of contemporary crash-frequency models.

The remainder of the paper is organized as follows. Section~\ref{sec:data} describes the networks and crash data. Sections~\ref{sec:model}--\ref{sec:computation} present the proposed model, its comparators, and posterior computation. Section~\ref{sec:simulation} reports the simulation study, and Sections~\ref{sec:barcelona-results}--\ref{sec:predictive-comparison} present the applications and predictive comparisons. Sections~\ref{sec:discussion} and~\ref{sec:conclusions} discuss the findings and conclude. The appendices provide additional spectral visualizations and representative simulated fields.

\section{Urban Road Networks and Crash Data}
\label{sec:data}

\subsection{Barcelona network construction}
\label{sec:barcelona-network}

The Barcelona road-network object is represented in the ETRS89/UTM zone 31N coordinate reference system and contains 2,714 road segments and 2,304 intersections distributed across 103 connected components. Spectral modeling is restricted to the largest connected component, comprising 2,540 segments and 2,028 intersections. This restriction excludes 174 segments (6.4\% of the available segments) and 276 intersections; inference therefore pertains to the retained component rather than the entire available network. Crash data were obtained from the Barcelona City Council Open Data portal (Open Data BCN; \url{https://opendata-ajuntament.barcelona.cat/en}) and cover 2010--2015.

\subsection{Crash counts and provisional exposure}
\label{sec:counts-exposure}

\begin{figure}[htbp]
  \centering
  \includegraphics[width=0.58\textwidth]{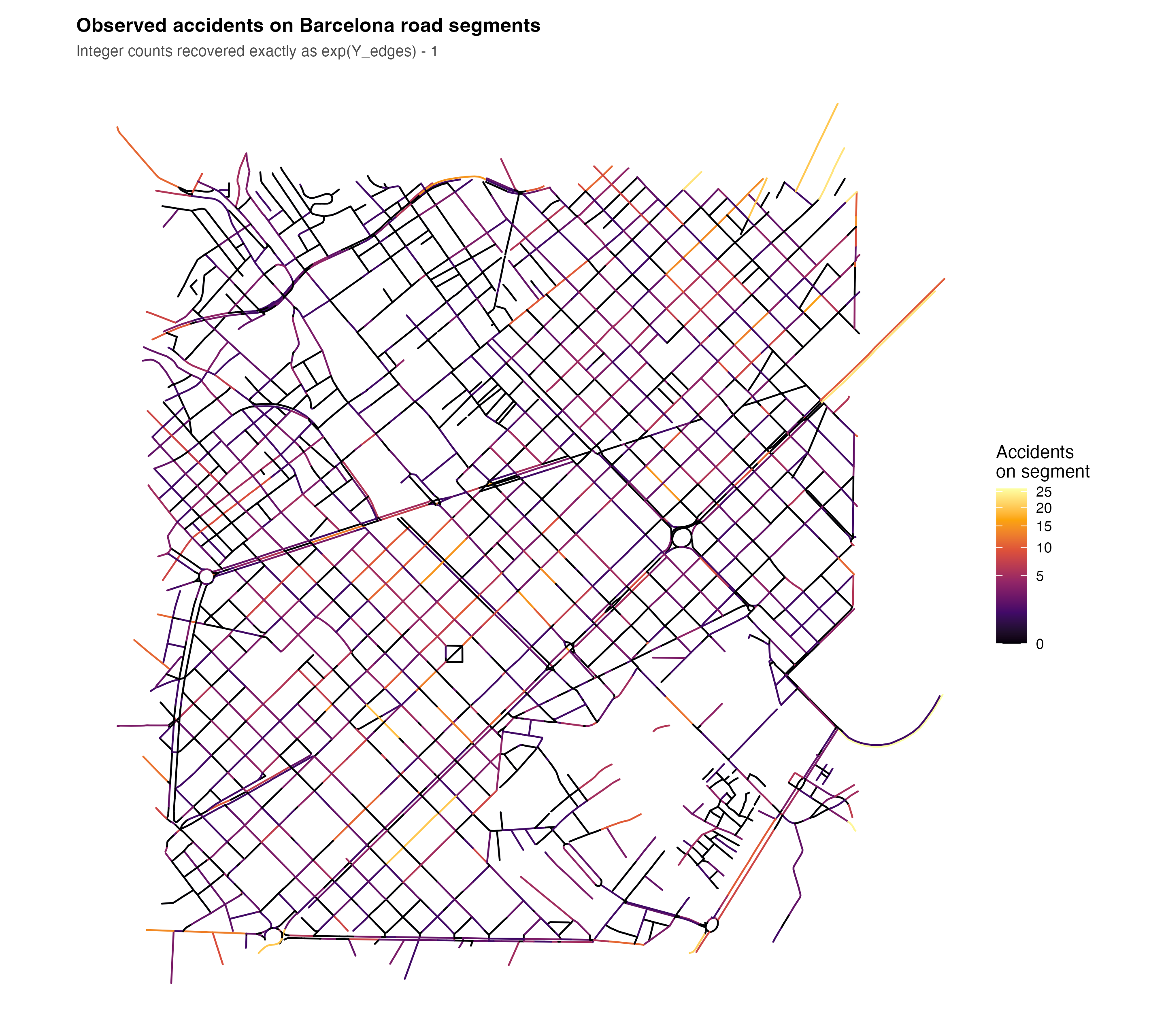}
  \caption{Recovered crash counts on Barcelona road segments, obtained by reversing the stored $\log(1+Y_e)$ transformation. Colors indicate the recovered counts $Y_e$.}
  \label{fig:barcelona-crash-counts}
\end{figure}

The stored segment-level response was verified to be numerically consistent with $\log(1+Y_e)$ for a nonnegative integer count $Y_e$. Reversing this transformation yields a total of 3,880 crashes across the full set of 2,714 segments, of which 1,501 (55.3\%) have zero counts. The mean and variance of the recovered counts are 1.43 and 7.56, respectively. Their discrepancy indicates marginal overdispersion, although it does not by itself establish overdispersion conditional on exposure and latent spatial effects. These descriptive summaries refer to the full network, whereas model fitting uses only its largest connected component. Compatibility with an integer-valued transformation is a numerical consistency check and does not independently validate the original event records, their deduplication, or their assignment to segments. Figure~\ref{fig:barcelona-crash-counts} displays the recovered segment-level counts.

The stored intersection-level response is a derived quantity equal to the sum of transformed responses over incident segments, rather than an independently observed crash count. It is therefore excluded from the likelihood, whose observational unit is the road segment. Segment lengths are calculated from the projected line geometries. Across the full network, the represented road length is 245.0~km, with mean and median segment lengths of 90.3~m and 64.8~m, respectively. The logarithm of segment length enters the model as an offset, imposing proportionality between expected counts and length when other model components are held fixed. Length serves as a provisional exposure proxy: the resulting rates describe crash frequency per unit road length over the observation period, rather than traffic-volume-adjusted crash risk.

\subsection{Bogot\'a network and data audit}
\label{sec:bogota-network}

The Bogot\'a analysis focuses on a central study area comprising eight localities: Chapinero, Teusaquillo, Barrios Unidos, Los M\'artires, La Candelaria, Santa Fe, Antonio Nari\~no, and Puente Aranda. Segments whose midpoints fall within the union of these localities form an initial set of 8,165 candidate segments. Spectral modeling uses its largest connected component, containing 6,076 segments and 5,104 intersections. This restriction excludes 2,089 candidate segments (25.6\%), whose aggregate crash count and total length are not documented in the available summaries. Inference is consequently limited to the retained component, and sensitivity to the exclusion of smaller components has not been assessed.

Crash records were obtained from the official Bogot\'a Open Data portal through the \textit{Siniestros Viales Consolidados Bogot\'a D.C.} dataset (\url{https://datosabiertos.bogota.gov.co/dataset/siniestros-viales-consolidados-bogota-d-c}), which contains traffic-incident records documented in police reports from 2015 onward. This source coverage does not identify the exact observation period represented by the analyzed network object, which remains to be documented. As in Barcelona, the stored segment-level values are numerically consistent with $\log(1+Y_e)$ for nonnegative integer counts. Reversing the transformation yields 85,021 crashes across the modeled component, including 1,722 zero-count segments (28.3\%). The mean count is 13.99 and its variance is 899.62, indicating substantial marginal heterogeneity. These numerical checks do not independently establish event-level validity or the accuracy of segment assignment.

\begin{figure}[htbp]
  \centering
  \includegraphics[width=0.52\textwidth]{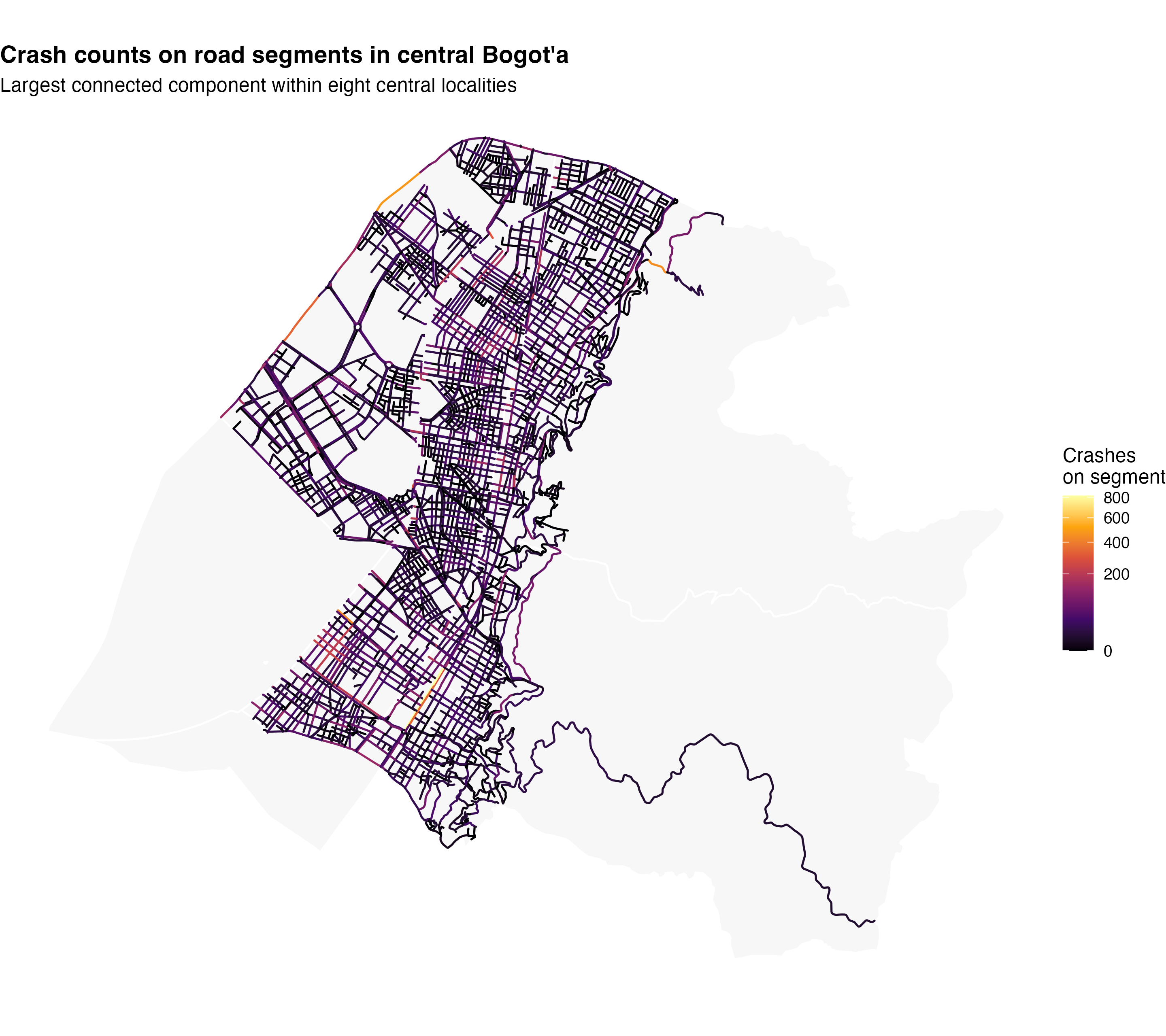}
  \caption{Recovered crash counts on the largest connected road component within the central Bogot\'a study area. Segments were selected by midpoint membership in the union of the eight study localities. Colors indicate counts obtained by reversing the stored $\log(1+Y_e)$ transformation.}
  \label{fig:bogota-crash-counts}
\end{figure}

The retained network represents 778.9~km of road, with mean and median segment lengths of 128.2~m and 91.5~m, respectively. Figure~\ref{fig:bogota-crash-counts} displays the recovered counts on this component. Segment length is used as an offset with the same interpretation as in Barcelona. However, the Barcelona descriptive summaries refer to the full available network, whereas the Bogot\'a summaries refer to the modeled component. Differences in spatial coverage, observation periods, and potentially event definitions preclude interpreting the reported count summaries as direct evidence of differences in crash risk between the two cities.

\section{Bayesian Edge Spectral Crash Model}
\label{sec:model}

\subsection{Road graph, observational support, and edge neighborhoods}
\label{sec:edge-neighborhoods}

Let $G=(V,E)$ be an undirected, loop-free road graph with $n=|V|$ intersections and $p=|E|$ road segments. The unsigned incidence matrix $\mathbf{C}\in\{0,1\}^{n\times p}$ has entry $C_{ve}=1$ if intersection $v$ is an endpoint of segment $e$, and zero otherwise. Each column therefore contains exactly two ones. The observed count vector $\boldsymbol{Y}=(Y_1,\ldots,Y_p)^{\top}$ and latent spatial field $\boldsymbol{s}=(s_1,\ldots,s_p)^{\top}$ are both indexed by road segments. Intersections determine neighborhood relationships, but the likelihood acts directly on the segment-level effects; no projection of intersection-level effects is required.

RENeGe represents spatial dependence through shared endpoints \citep{cruzreyes2023spasta}. Its associated line graph $L(G)$ has one vertex for each road segment, with adjacency defined by
\begin{equation}
  (\mathbf{A}_E)_{e\ell}
  =
  \textsf{I}(e\neq\ell)\,
  \textsf{I}\!\big((\mathbf{C}^{\top}\mathbf{C})_{e\ell}>0\big),
  \label{eq:edge-adjacency}
\end{equation}
where $\textsf{I}(\cdot)$ denotes the indicator function. For a simple road graph, this reduces to $\mathbf{A}_E=\mathbf{C}^{\top}\mathbf{C}-2\mathbf{I}_p$. If parallel segments share both endpoints, however, the latter expression assigns weight two to their relationship rather than binary adjacency. Equation~\eqref{eq:edge-adjacency} defines the binary construction in either case. The available network summaries do not establish whether the fitted operators used simple graphs or explicitly binarized adjacency; this preprocessing detail must therefore be verified before identifying the implemented operator with the binary line graph.

Choosing edge rather than node support changes the class of spatial patterns that can be represented. Mapping a nodal field to segments through endpoint averages constrains the segment field to combinations of shared intersection effects. An edge-supported field instead assigns a separate value to each segment, although adjacency-based regularization still couples neighboring values and spectral truncation restricts the available patterns. This construction can accommodate contrasts among incident segments, but it does not imply universally greater spatial resolution or less smoothing than a nodal model. Those properties depend jointly on the operator, retained basis, and prior.

Intersection-level summaries can nevertheless be derived after fitting. Let $\mathbf{D}_I=\textsf{diag}\!\big(\mathbf{C}\boldsymbol{1}_p\big)$ contain the numbers of modeled segments incident on each retained intersection, assumed to be positive. For posterior draw $b$, define
\begin{equation}
  \boldsymbol{r}^{(b)}
  =
  \mathbf{D}_I^{-1}\mathbf{C}\boldsymbol{s}^{(b)}.
  \label{eq:posterior-edge-node-map}
\end{equation}
Each entry is the average latent effect over segments incident on an intersection. Applying this transformation to every draw propagates posterior uncertainty, and linearity gives $\textsf{E}\big(\boldsymbol{r}\mid\boldsymbol{Y}\big)=\mathbf{D}_I^{-1}\mathbf{C}\,\textsf{E}\big(\boldsymbol{s}\mid\boldsymbol{Y}\big)$. This is a descriptive projection of the fitted segment field, not a separate intersection-level model or an aggregation of expected crash counts.

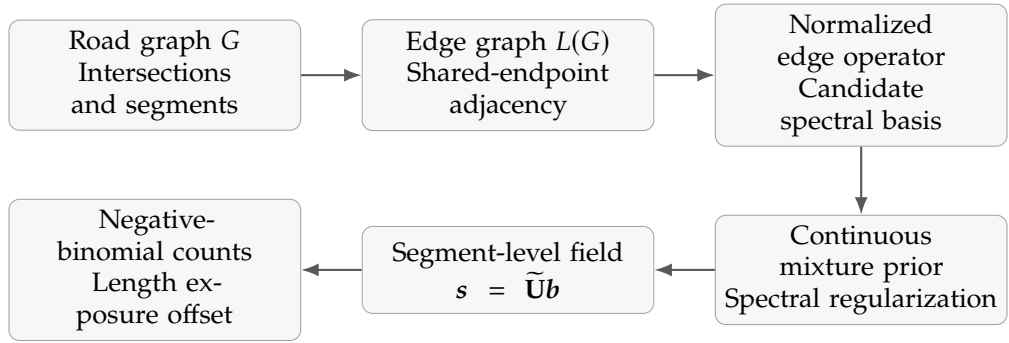
\begin{figure}[htbp]
  \centering
  \begin{tikzpicture}[
    block/.style={
      draw=black!35,
      rounded corners,
      fill=black!3,
      align=center,
      text width=3.6cm,
      minimum height=1.15cm,
      font=\small
    },
    arr/.style={-{Latex},thick,draw=black!65},
    node distance=9mm and 8mm
  ]
    \node[block] (road)
      {Road graph $G$\\Intersections and segments};
    \node[block,right=of road] (edge)
      {Edge graph $L(G)$\\Shared-endpoint adjacency};
    \node[block,right=of edge] (basis)
      {Normalized edge operator\\Candidate spectral basis};
    \node[block,below=of basis] (prior)
      {Continuous mixture prior\\Spectral regularization};
    \node[block,left=of prior] (field)
      {Segment-level field\\$\boldsymbol{s}=\widetilde{\mathbf{U}}\boldsymbol{b}$};
    \node[block,left=of field] (counts)
      {Negative-binomial counts\\Length exposure offset};

    \draw[arr] (road) -- (edge);
    \draw[arr] (edge) -- (basis);
    \draw[arr] (basis) -- (prior);
    \draw[arr] (prior) -- (field);
    \draw[arr] (field) -- (counts);
  \end{tikzpicture}
  \caption{Construction of the proposed model. The edge-neighborhood operator supplies candidate segment-level patterns, whose coefficients receive continuous spike-and-slab regularization. The resulting spatial field enters a negative-binomial model for crash counts.}
  \label{fig:model-construction}
\end{figure}

\subsection{RENeGe precision and spectral representation}
\label{sec:renege-representation}

Let $\mathbf{M}_E=\textsf{diag}\!\big(\mathbf{A}_E\boldsymbol{1}_p\big)$ be the edge-degree matrix. On a connected edge graph with at least two vertices, all diagonal entries are positive. A proper RENeGe-type precision matrix is
\begin{equation}
  \mathbf{Q}_E(\gamma)
  =
  \kappa\big(\mathbf{M}_E-\gamma\mathbf{A}_E\big),
  \qquad
  0<\gamma<1,\quad \kappa>0,
  \label{eq:renege-precision}
\end{equation}
where $\gamma$ controls spatial dependence and $\kappa$ is a global precision multiplier. Marginal variances depend on both parameters and the network geometry, so $\kappa^{-1}$ is not itself a common marginal variance. Defining the symmetric normalized adjacency operator as $\mathbf{S}_E=\mathbf{M}_E^{-1/2}\mathbf{A}_E\mathbf{M}_E^{-1/2}$ gives
\begin{equation}
  \mathbf{Q}_E(\gamma)^{-1}
  =
  \kappa^{-1}\mathbf{M}_E^{-1/2}
  \big(\mathbf{I}_p-\gamma\mathbf{S}_E\big)^{-1}
  \mathbf{M}_E^{-1/2}.
  \label{eq:renege-covariance}
\end{equation}
The eigenvalues of $\mathbf{S}_E$ lie in $[-1,1]$, ensuring positive definiteness under the stated restriction on $\gamma$.

Write $\mathbf{S}_E=\mathbf{U}\boldsymbol{\Delta}\mathbf{U}^{\top}$, where $\boldsymbol{\Delta}$ is here replaced by the matrix notation $\mathbf{\Delta}=\textsf{diag}(\delta_1,\ldots,\delta_p)$, so that, more precisely, $\mathbf{S}_E=\mathbf{U}\mathbf{\Delta}\mathbf{U}^{\top}$. The columns of $\mathbf{U}=(\boldsymbol{u}_1,\ldots,\boldsymbol{u}_p)$ are orthonormal, and the eigenvalues satisfy $1=\delta_1\geq\delta_2\geq\cdots\geq\delta_p\geq-1$. The covariance becomes
\begin{equation}
  \mathbf{Q}_E(\gamma)^{-1}
  =
  \kappa^{-1}\mathbf{M}_E^{-1/2}
  \mathbf{U}\,
  \textsf{diag}\!\big(
    (1-\gamma\delta_1)^{-1},\ldots,
    (1-\gamma\delta_p)^{-1}
  \big)
  \mathbf{U}^{\top}\mathbf{M}_E^{-1/2}.
  \label{eq:spectral-covariance}
\end{equation}
This representation motivates eigenvalue-dependent regularization. For $\gamma>0$, larger eigenvalues receive larger variance weights. They also correspond to smaller eigenvalues of the normalized Laplacian $\mathbf{I}_p-\mathbf{S}_E$ and hence lower normalized graph variation. This notion of smoothness is defined by the graph operator, rather than by geographic distance.

The leading eigenvector satisfies $\boldsymbol{u}_1\propto\mathbf{M}_E^{1/2}\boldsymbol{1}_p$ and is generally nonconstant, although its degree-adjusted counterpart $\mathbf{M}_E^{-1/2}\boldsymbol{u}_1$ is constant. The fitted model excludes $\boldsymbol{u}_1$ and retains $\boldsymbol{u}_2,\ldots,\boldsymbol{u}_{M+1}$. Because the fitted basis uses the eigenvectors without degree adjustment, exclusion of the leading vector is a modeling restriction and is not fully justified by the presence of an intercept; it may remove degree-related variation.

For candidate $j=1,\ldots,M$, let $\boldsymbol{v}_j=\boldsymbol{u}_{j+1}$ and $\lambda_j=\delta_{j+1}$. All reported mode numbers refer to this candidate index. Define $\bar v_j=p^{-1}\sum_{e=1}^{p}v_{ej}$ and the positive root-mean-square scale $d_j=\{p^{-1}\sum_{e=1}^{p}(v_{ej}-\bar v_j)^2\}^{1/2}$. The fitted basis has entries
\begin{equation}
  \widetilde u_{ej}
  =
  \frac{v_{ej}-\bar v_j}{d_j},
  \qquad
  \widetilde{\mathbf{U}}
  =
  \mathbf{P}\mathbf{U}_c\mathbf{D}^{-1},
  \label{eq:rms-standardization}
\end{equation}
where $\mathbf{P}=\mathbf{I}_p-p^{-1}\boldsymbol{1}_p\boldsymbol{1}_p^{\top}$, $\mathbf{U}_c=(\boldsymbol{u}_2,\ldots,\boldsymbol{u}_{M+1})$, and $\mathbf{D}=\textsf{diag}(d_1,\ldots,d_M)$. Centering separates the spatial field from the intercept by imposing $\boldsymbol{1}_p^{\top}\boldsymbol{s}=0$, while scaling places the candidate columns on a common empirical scale. Centering does not generally preserve their eigenvector property or mutual orthogonality, so the $\lambda_j$ remain eigenvalues associated with the original, untransformed vectors.

Under the coefficient prior introduced below, assigning all candidates to the slab gives
\begin{equation}
\begin{split}
  \textsf{Var}\big(
    \boldsymbol{s}\mid\tau,\gamma,\boldsymbol{z}=\boldsymbol{1}_M
  \big)
  ={}&
  \tau^2\mathbf{P}\mathbf{U}_c\mathbf{D}^{-1}
  \textsf{diag}\!\big(
    (1-\gamma\lambda_1)^{-1},\ldots,
    (1-\gamma\lambda_M)^{-1}
  \big)\\
  &{}\times
  \mathbf{D}^{-1}\mathbf{U}_c^{\top}\mathbf{P}.
\end{split}
\label{eq:induced-spectral-covariance}
\end{equation}
This differs from a truncation of Eq.~\eqref{eq:spectral-covariance}: the fitted basis omits the degree factors $\mathbf{M}_E^{-1/2}$ and introduces centering and scaling. The proposed model is therefore an operator-inspired spectral construction rather than an exact reparameterization of the RENeGe Gaussian field. Eigenvectors of a projected operator such as $\mathbf{P}_X\mathbf{S}_E\mathbf{P}_X$, restricted to the range of $\mathbf{P}_X=\mathbf{I}_p-\mathbf{X}(\mathbf{X}^{\top}\mathbf{X})^{-1}\mathbf{X}^{\top}$ for a full-column-rank design matrix $\mathbf{X}$, would define a different basis and were not used here.

Eigenvector signs are arbitrary, and repeated eigenvalues additionally permit rotations within their eigenspaces. Sign reversals leave fitted contributions unchanged when accompanied by corresponding coefficient reversals. Component-specific selection, however, can depend on the basis chosen within a repeated eigenspace. Accordingly, individual-mode summaries are conditional on the computed candidate basis and should be interpreted through their fitted spatial contributions rather than through eigenvector colors or signs alone.

\subsection{Negative-binomial observation model}
\label{sec:observation-model}

For segment $e$, let $Y_e$ denote the crash count, $L_e>0$ its length in kilometres, and $\boldsymbol{x}_e$ a column vector of measured road characteristics, excluding the intercept. Conditional on the latent field and model parameters, counts are assumed independent, with
\begin{equation}
\begin{aligned}
  Y_e\mid\mu_e,\phi
  &\sim \textsf{NB}_2(\mu_e,\phi),\\
  \textsf{log}(\mu_e)
  &=
  \textsf{log}(L_e)+\alpha_0
  +\boldsymbol{x}_e^{\top}\boldsymbol{\alpha}+s_e,
  \qquad
  \boldsymbol{s}=\widetilde{\mathbf{U}}\boldsymbol{b}.
\end{aligned}
\label{eq:observation-model}
\end{equation}
Here, $\textsf{E}\big(Y_e\mid\mu_e,\phi\big)=\mu_e$ and $\textsf{Var}\big(Y_e\mid\mu_e,\phi\big)=\mu_e+\mu_e^2/\phi$, with $\phi>0$. The offset fixes the coefficient of log length at one, imposing proportionality between expected counts and segment length conditional on the remaining terms. Thus, $\textsf{exp}(\alpha_0+\boldsymbol{x}_e^{\top}\boldsymbol{\alpha}+s_e)$ represents expected crash frequency per kilometre over the observation period. This proportionality assumption has not been assessed by estimating the log-length coefficient, and length does not account for traffic volume. For a model including measured road-level covariates, we assign independent weakly informative priors

\[
\alpha_k \sim N(0,1), \qquad k=1,\ldots,q,
\]

after standardizing continuous covariates.

The applications include only an intercept and spatial effects; the covariate term in Eq.~\eqref{eq:observation-model} describes an extension rather than a fitted component. Consequently, the spatial field may absorb unmeasured roadway and traffic heterogeneity. The negative-binomial likelihood respects the discrete, nonnegative support of the response and permits conditional overdispersion relative to a Poisson model. Nevertheless, marginal overdispersion alone does not establish the need for this additional conditional variation, because heterogeneous exposures and latent effects can also increase marginal variance. Similarly, a large proportion of zeros does not by itself establish zero inflation. Posterior predictive checks of zeros, upper tails, and dispersion are needed to assess these aspects of fit and are not available in the reported analysis.

\subsection{Spectral spike-and-slab prior}
\label{sec:spike-slab}

Let $z_j\in\{0,1\}$ indicate membership in the component designated as the slab. Conditional on the hyperparameters, the indicators are independent and the coefficients follow
\begin{equation}
\begin{aligned}
  z_j\mid\pi
  &\sim \textsf{Bernoulli}(\pi),\\
  b_j\mid z_j,\tau,\gamma
  &\sim
  \begin{cases}
    \textsf{N}(0,\sigma_0^2), & z_j=0,\\[2pt]
    \textsf{N}\!\big(0,\tau^2/(1-\gamma\lambda_j)\big), & z_j=1.
  \end{cases}
\end{aligned}
\label{eq:spike-slab-prior}
\end{equation}
The normal distributions are parameterized by their means and variances, and $\sigma_0>0$ is fixed. Computation integrates out the indicators, yielding the continuous mixture $b_j\mid\pi,\tau,\gamma\sim(1-\pi)\textsf{N}(0,\sigma_0^2)+\pi\textsf{N}\!\big(0,\tau^2/(1-\gamma\lambda_j)\big)$. Both components have continuous support, so coefficients are not set exactly to zero. Throughout, inclusion refers to slab membership rather than a nonzero coefficient.

The eigenvalue-dependent slab variance follows the ordering in Eq.~\eqref{eq:spectral-covariance}, assigning larger prior variances to candidates associated with larger eigenvalues. This ordering motivates regularization but does not restore the exact covariance or smoothness properties altered by basis transformation. Moreover, the prior on $\tau$ does not guarantee that the slab is wider than the spike. The inequality $\tau/\sqrt{1-\gamma\lambda_j}>\sigma_0$ holds for every candidate only when $\tau>\sigma_0\max_{1\leq j\leq M}\sqrt{1-\gamma\lambda_j}$. Its posterior probability cannot be recovered from the available summaries. Slab membership therefore denotes a mixture label whose interpretation as a relatively weakly shrunk contribution requires verification; imposing ordered scales would change the fitted prior.

We assign $\pi\sim\textsf{Beta}(1,4)$ to favor a small proportion of slab memberships without fixing their number. With $K=\sum_{j=1}^{M}z_j$, the conditional prior is $K\mid\pi\sim\textsf{Binomial}(M,\pi)$, and integrating out $\pi$ gives
\begin{equation}
  \textsf{P}(K=k)
  =
  \binom{M}{k}
  \frac{\textsf{B}(1+k,\,4+M-k)}{\textsf{B}(1,4)},
  \qquad k=0,\ldots,M,
  \label{eq:complexity-prior}
\end{equation}
where $\textsf{B}(\cdot,\cdot)$ is the beta function. Thus, $\textsf{E}(K)=M/5$, equal to four for the fitted candidate size $M=20$. This prior concerns mixture memberships, not the algebraic rank of the spatial field or the number of exactly nonzero coefficients.

The reported fits use $M=20$, $\gamma=0.9$, and $\sigma_0=0.05$, together with $\tau\sim\textsf{N}^{+}(0,0.5^2)$ and $\phi\sim\textsf{Exponential}(0.5)$, where $\textsf{N}^{+}$ denotes a normal distribution truncated to the positive half-line and the exponential parameter is a rate. The intercept has a normal prior with unit variance centered at the empirical log crash frequency per kilometre. This data-dependent centering is an empirical-Bayes choice: its value changes between datasets, and retaining the full-data center during leave-one-out evaluation allows held-out observations to influence the prior. The resulting predictive scores are therefore exploratory assessments under the fitted specification, rather than validation using a prior specified independently of the evaluation data. A fully training-based assessment would require recomputing the prior center within each training set.

The restriction to 20 candidates is a computational truncation, not an empirically established optimum. Sensitivity to the candidate size, dependence parameter, spike scale, inclusion prior, and slab-scale prior has not been evaluated. These choices jointly determine which patterns can be represented and how mixture membership is interpreted; posterior component summaries are conditional on this specification.

\subsection{Posterior estimands and interpretation}
\label{sec:estimands}

The segment-level field is $s_e=\sum_{j=1}^{M}b_j\widetilde u_{ej}$, and the posterior mean contribution of candidate $j$ is $\boldsymbol{a}_j=\widetilde{\boldsymbol{u}}_j\,\textsf{E}\big(b_j\mid\boldsymbol{y}\big)$. These contributions sum to $\textsf{E}\big(\boldsymbol{s}\mid\boldsymbol{y}\big)$ and remain unchanged under simultaneous sign reversal of a basis column and its coefficient. Posterior slab-membership probabilities $\textsf{P}\big(z_j=1\mid\boldsymbol{y}\big)$ summarize component allocation, while their sum gives $\textsf{E}\big(K\mid\boldsymbol{y}\big)=\sum_{j=1}^{M}\textsf{P}\big(z_j=1\mid\boldsymbol{y}\big)$. This expected membership count provides a model-dependent summary of spectral complexity, subject to the scale-ordering qualification above. It does not measure effective degrees of freedom, and its mean alone does not describe posterior uncertainty in $K$.

Expected crash counts require averaging over the joint posterior:
\begin{equation}
  \textsf{E}\big(\mu_e\mid\boldsymbol{y}\big)
  =
  L_e\,\textsf{E}\!\left(
    \textsf{exp}\!\big(
      \alpha_0+\boldsymbol{x}_e^{\top}\boldsymbol{\alpha}+s_e
    \big)
    \,\middle|\,\boldsymbol{y}
  \right).
  \label{eq:posterior-mean-count}
\end{equation}
For posterior draws $b=1,\ldots,B$, this expectation is estimated by $B^{-1}\sum_{b=1}^{B}\mu_e^{(b)}$, where $\mu_e^{(b)}=L_e\,\textsf{exp}\!\big(\alpha_0^{(b)}+\boldsymbol{x}_e^{\top}\boldsymbol{\alpha}^{(b)}+s_e^{(b)}\big)$. All candidate coefficients contribute to each draw; thresholding slab-membership probabilities is unnecessary and generally changes the posterior summary. Dividing by $L_e$ gives posterior mean frequency per kilometre. Credible intervals for expected counts or frequencies can likewise be obtained from the corresponding draws, whereas predictive intervals for future counts must additionally incorporate negative-binomial sampling variation.

The reported frequency maps instead use the plug-in quantity $\widehat f_e^{\textsf{plug}}=\textsf{exp}\!\big(\textsf{E}(\alpha_0\mid\boldsymbol{y})+\boldsymbol{x}_e^{\top}\textsf{E}(\boldsymbol{\alpha}\mid\boldsymbol{y})+\textsf{E}(s_e\mid\boldsymbol{y})\big)$. Because exponentiation is nonlinear, this generally differs from posterior mean frequency per kilometre. These maps describe fitted frequency evaluated at posterior mean parameters and do not display posterior uncertainty; contribution maps separately summarize the linear log-frequency field. Predictive counts, exceedance probabilities, Monte Carlo standard errors for slab-membership probabilities, and the posterior distribution of $K$ are not reported.

\section{Spectral CAR and BYM2 Comparison Models}
\label{sec:comparison-models}
\label{sec:spectral-comparators}

All spatial models are defined on the same road-segment support and share the negative-binomial likelihood, log-length offset, edge-neighborhood operator, and centered, RMS-standardized candidate basis $\widetilde{\mathbf{U}}$. The comparisons assess alternative regularization and heterogeneity structures within this common representation. Sparse RENeGe assigns continuous spike-and-slab priors to the spectral coefficients, spectral CAR assigns Gaussian priors without mixture indicators, and spectral BYM2 supplements the Gaussian structured field with independent segment-level effects. Since the candidate columns are transformed eigenvectors, the CAR and BYM2 labels describe spectral analogues rather than exact representations of their full-rank counterparts.

\subsection{Spectral CAR on the RENeGe basis}
\label{sec:spectral-car}
\label{sec:edge-car-spectral}

The covariance representation in Eq.~\eqref{eq:spectral-covariance} motivates the weights $w_j=(1-\gamma\lambda_j)^{-1/2}$ for candidates $j=1,\ldots,M$. Using the same standardized columns as Sparse RENeGe, define
\begin{equation}
  g_e
  =
  \frac{\sum_{j=1}^{M}w_j\xi_j\widetilde u_{ej}}
       {\big(\sum_{j=1}^{M}w_j^2\big)^{1/2}},
  \qquad
  \xi_j\overset{\textsf{iid}}{\sim}\textsf{N}(0,1).
  \label{eq:edge-car-common-basis}
\end{equation}
The resulting field $\boldsymbol{g}=(g_1,\ldots,g_p)^{\top}$ has zero prior mean and satisfies $\boldsymbol{1}_p^{\top}\boldsymbol{g}=0$. Since each candidate column has unit mean square and the coefficients are independent, its average marginal variance is $p^{-1}\sum_{e=1}^{p}\textsf{Var}(g_e)=1$. This normalization does not require orthogonality among the transformed columns and does not imply equal marginal variances across segments.

The spectral CAR predictor is
\begin{equation}
  \textsf{log}(\mu_e)
  =
  \textsf{log}(L_e)+\alpha_0+\sigma_s g_e,
  \qquad \sigma_s>0.
  \label{eq:spectral-car}
\end{equation}
Thus, $\sigma_s^2$ is the average marginal variance of the latent structured effect, conditional on $\sigma_s$. All $M$ candidate coefficients receive Gaussian regularization, without mixture-based membership probabilities. Equivalently, the coefficient of $\widetilde{\boldsymbol{u}}_j$ has conditional variance $\sigma_s^2w_j^2/\sum_{\ell=1}^{M}w_\ell^2$. The eigenvalue-dependent variance ratios match those of the Sparse RENeGe slab, but their overall scaling and hyperprior specification need not coincide. Consequently, this comparison evaluates the fitted regularization schemes within a common basis rather than isolating the effect of component selection alone. As discussed in Section~\ref{sec:renege-representation}, the transformed basis also means that this field is not an exact truncation of the proper edge-CAR covariance.

\subsection{Spectral BYM2 on the RENeGe basis}
\label{sec:spectral-bym2}
\label{sec:edge-bym2}

The spectral BYM2 model combines the structured field in Eq.~\eqref{eq:edge-car-common-basis} with independent segment-level Gaussian effects:
\begin{equation}
\begin{aligned}
  \textsf{log}(\mu_e)
  &=
  \textsf{log}(L_e)+\alpha_0+
  \sigma\big\{\sqrt{\rho}\,g_e+\sqrt{1-\rho}\,v_e\big\},\\
  v_e
  &\overset{\textsf{iid}}{\sim}\textsf{N}(0,1),
  \qquad \sigma>0,\quad 0<\rho<1,
\end{aligned}
\label{eq:spectral-bym2}
\end{equation}
where $\boldsymbol{v}=(v_1,\ldots,v_p)^{\top}$ is independent of $\boldsymbol{g}$. The structured component is low rank, whereas the independent component permits variation outside the retained spectral span. This construction is a spectral BYM2 analogue with a proper, truncated structured field.

Let $h_e=\sigma\{\sqrt{\rho}\,g_e+\sqrt{1-\rho}\,v_e\}$. Conditional on the hyperparameters, its variance is $\textsf{Var}(h_e\mid\sigma,\rho)=\sigma^2\{\rho\,\textsf{Var}(g_e)+(1-\rho)\}$, so $p^{-1}\sum_{e=1}^{p}\textsf{Var}(h_e\mid\sigma,\rho)=\sigma^2$. Accordingly, $\sigma$ controls the overall latent scale, and $\rho$ is the ratio of average structured variance to average total latent variance. It is not generally the structured variance fraction at an individual segment, which also depends on $\textsf{Var}(g_e)$. Values near one favor structured variation within the retained basis; values near zero favor independent segment-level heterogeneity.

The negative-binomial dispersion parameter $\phi$ provides a separate source of conditional count variation. Therefore, $\rho$ does not represent the proportion of total count heterogeneity explained by network structure. Both negative-binomial dispersion and independent latent effects can accommodate excess variation, potentially weakening their separation in the posterior. The extent of this overlap has not been assessed in the reported fits.

\subsection{Comparison logic}
\label{sec:comparison-logic}
\label{sec:spatial-model-summary}

Table~\ref{tab:spatial-model-summary} summarizes the three spatial formulations. Sparse RENeGe and spectral CAR share the candidate basis but differ in coefficient regularization, scale specification, and the availability of posterior mixture-membership summaries. Spectral CAR and spectral BYM2 share the normalized structured field, with the latter allowing additional independent segment-level variation. These comparisons are conditional on the retained basis and fitted priors; they do not establish the relative performance of full-rank spatial models or alternative basis constructions.

\begin{table}[htbp]
  \centering
  \caption{Spatial formulations using the common centered and RMS-standardized edge spectral basis. All candidates remain in the continuous Sparse RENeGe representation; slab membership does not imply exact exclusion of the remaining coefficients.}
  \label{tab:spatial-model-summary}
  \small
  \renewcommand{\arraystretch}{1.2}
  \begin{tabularx}{\textwidth}{
    @{}
    >{\raggedright\arraybackslash}p{2.5cm}
    >{\raggedright\arraybackslash}X
    >{\raggedright\arraybackslash}p{2.8cm}
    >{\raggedright\arraybackslash}p{2.7cm}
    @{}
  }
    \toprule
    \textbf{Model} &
    \textbf{Structured component} &
    \textbf{Component assessment} &
    \textbf{Independent latent effects} \\
    \midrule
    Sparse RENeGe &
    Continuous spike-and-slab regularization of all candidate coefficients &
    Posterior slab-membership probabilities &
    None \\
    \addlinespace
    Spectral CAR &
    Gaussian regularization of all candidate coefficients &
    No mixture indicators &
    None \\
    \addlinespace
    Spectral BYM2 &
    Same normalized Gaussian field as spectral CAR &
    No mixture indicators &
    Segment-specific Gaussian effects \\
    \bottomrule
  \end{tabularx}
\end{table}

A nonspatial negative-binomial model with predictor $\textsf{log}(\mu_e)=\textsf{log}(L_e)+\alpha_0$ provides a baseline for assessing the predictive contribution of latent heterogeneity beyond the length offset and negative-binomial dispersion. Improvements over this baseline do not, by themselves, isolate the contribution of network dependence, particularly for BYM2, which also includes independent effects. Interpretation of all comparisons additionally depends on the priors assigned to $\sigma_s$, $\sigma$, and $\rho$, as well as on consistent treatment of the shared intercept and dispersion parameters.

\section{Posterior Computation}
\label{sec:computation}

Posterior inference uses the $M=20$ candidate basis functions defined in Section~\ref{sec:renege-representation}. The discrete mixture indicators are integrated out analytically, allowing the continuous parameters to be sampled in Stan using the No-U-Turn Sampler (NUTS), an adaptive Hamiltonian Monte Carlo algorithm. For candidate $j$, the marginalized prior density is
\begin{equation}
  p(b_j\mid\pi,\tau,\gamma)
  =
  \pi f_{\textsf{N}}(b_j;0,\sigma_{1j}^2)
  +(1-\pi)f_{\textsf{N}}(b_j;0,\sigma_0^2),
  \qquad
  \sigma_{1j}=\frac{\tau}{\sqrt{1-\gamma\lambda_j}},
  \label{eq:marginalized-prior}
\end{equation}
where $f_{\textsf{N}}(\cdot;0,\sigma^2)$ denotes a normal density with mean zero and variance $\sigma^2$, and $\sigma_0=0.05$. Marginalization preserves the specified continuous-mixture model without introducing discrete sampling steps; it does not set coefficients exactly to zero.

For retained posterior draw $b$, the conditional slab-membership probability is recovered by Bayes' rule:
\begin{equation}
  q_j^{(b)}
  =
  \frac{
    \pi^{(b)}
    f_{\textsf{N}}\!\big(
      b_j^{(b)};0,(\sigma_{1j}^{(b)})^2
    \big)
  }{
    \pi^{(b)}
    f_{\textsf{N}}\!\big(
      b_j^{(b)};0,(\sigma_{1j}^{(b)})^2
    \big)
    +(1-\pi^{(b)})
    f_{\textsf{N}}\!\big(b_j^{(b)};0,\sigma_0^2\big)
  }.
  \label{eq:conditional-slab-probability}
\end{equation}
The posterior inclusion probability, understood here as slab membership, is $\textsf{P}\big(z_j=1\mid\boldsymbol{y}\big)=\textsf{E}\big(q_j\mid\boldsymbol{y}\big)$ and is estimated by $B^{-1}\sum_{b=1}^{B}q_j^{(b)}$. This averages conditional probabilities rather than sampled indicators. Quantiles of the $q_j^{(b)}$ describe posterior variation in the conditional membership probability; they are neither credible intervals for the integrated inclusion probability nor Monte Carlo error bars. Monte Carlo uncertainty in its estimated mean must instead account for autocorrelation in the retained sequence.

Conditional on a draw of the continuous parameters, the indicators are independent Bernoulli variables with probabilities $q_1^{(b)},\ldots,q_M^{(b)}$. Consequently, the conditional distribution of $K=\sum_{j=1}^{M}z_j$ is Poisson-binomial, and its marginal posterior is a mixture of these distributions over the continuous posterior. Its mean is estimated by $B^{-1}\sum_{b=1}^{B}\sum_{j=1}^{M}q_j^{(b)}$, but this quantity does not characterize its dispersion or shape. Although discrete indicators are absent from NUTS, posterior draws of $K$ could be generated afterward by sampling the conditional Bernoulli indicators. The reported analysis summarizes expected membership counts without reconstructing the full posterior distribution of $K$.

For each city, the proposed model was fitted using four chains, each with 2,000 warm-up transitions followed by 2,000 sampling transitions. Retaining every second post-warm-up draw yielded 4,000 posterior draws per city. The target acceptance probability was 0.99, and the maximum tree depth was 13. Convergence and sampling efficiency were assessed using split $\widehat R$, effective sample sizes, divergent transitions, and tree-depth saturation. These diagnostics assess posterior computation rather than the adequacy of the observation model.

\subsection{Comparator models and predictive assessment}
\label{sec:predictive-assessment}

Sparse RENeGe is compared with the spectral CAR and spectral BYM2 formulations in Section~\ref{sec:spectral-comparators} and an intercept-only nonspatial negative-binomial baseline. All four models use the same segments, counts, length offsets, and intercept and dispersion priors; the spatial models also share the candidate basis. The comparator scale priors are $\sigma_s\sim\textsf{N}^{+}(0,0.5^2)$ and $\sigma\sim\textsf{N}^{+}(0,0.5^2)$, and the spectral BYM2 mixing parameter has prior $\rho\sim\textsf{Beta}(0.5,0.5)$. This specification uses the structured--unstructured decomposition associated with BYM2 \citep{riebler2016}, with the spectral construction and scaling described above. Assigning the same numerical half-normal scale to these parameters and the Sparse RENeGe slab parameter does not imply identical priors on total latent variation.

Predictive performance is assessed using Pareto-smoothed importance-sampling leave-one-out cross-validation (PSIS-LOO) \citep{vehtari2017}. The target is prediction of a segment count after withholding that count while retaining the network geometry, exposure information, and counts on the remaining segments. For $\boldsymbol{y}_{-e}$ denoting all observed counts except $y_e$, the leave-one-out log predictive score is
\begin{equation}
  \widehat{\textsf{ELPD}}_{\textsf{LOO}}
  =
  \sum_{e=1}^{p}
  \textsf{log}\,p(y_e\mid\boldsymbol{y}_{-e}),
  \label{eq:loo-elpd}
\end{equation}
with PSIS used to approximate the leave-one-out predictive densities from full-data posterior draws. We used 1,000 evenly spaced retained draws, comprising 250 from each chain, and report ELPD, its estimated standard error, paired model differences and their standard errors, the effective parameter count $p_{\textsf{LOO}}$, and $\textsf{LOOIC}=-2\widehat{\textsf{ELPD}}_{\textsf{LOO}}$. Standard errors based on pointwise contributions should be interpreted cautiously because observations are network-dependent.

For spectral BYM2, the segment-specific effect $v_e$ was integrated against its $\textsf{N}(0,1)$ prior using 15-point Gaussian quadrature before computing the pointwise predictive contributions. Conditional on the shared parameters, an independent effect for a segment with its count withheld receives no information from the other observations. Integration therefore aligns the predictive calculation with this held-out target and avoids directly evaluating the count using an effect learned from that same count. Properly implemented conditional importance sampling can also target leave-one-out prediction, but simply evaluating conditional posterior likelihoods does not constitute that calculation. Accuracy of the quadrature approximation has not been assessed by increasing the number of nodes.

The comparisons remain exploratory because the intercept prior is centered using the full dataset, including the observation being evaluated. Moreover, pointwise withholding retains information from neighboring segments through the shared spectral field. It therefore assesses prediction within the observed network rather than transfer to an unobserved region or prediction when neighboring segments are withheld together. Spatially blocked validation has not been performed. Conditional WAIC and DIC summaries are omitted because they would not provide a directly comparable assessment unless their treatment of segment-specific effects and predictive targets were aligned.

No posterior predictive checks of zero frequencies, upper-tail counts, dispersion, or residual edge autocorrelation are reported. Accordingly, satisfactory sampling diagnostics and favorable pointwise predictive scores do not establish distributional adequacy or geographic generalizability.

\section{Simulation Study}
\label{sec:simulation}

We conducted a simulation study to assess recovery of the latent segment-level log-frequency field, identification of its generating spectral components, and prediction of new crash counts. The experiment uses the Barcelona and central Bogot\'a network geometries, four generating mechanisms, and three composite difficulty levels. Each of the resulting 24 scenarios was independently replicated 50 times, yielding 1,200 simulated datasets. Fitting four models to each dataset produced 4,800 Bayesian model fits. The mechanisms range from exactly sparse fields in the retained basis to fields containing dense, unstructured, or nonlinear spectral variation. They assess robustness within the specified constructions rather than exhaustively representing possible road-network processes.

\subsection{Road-network spectral representation}
\label{sec:simulation-representation}

For each network $G=(V,E)$, with $p=|E|$ road segments, the simulation uses the same $M=20$ candidate columns as the empirical analysis. Let $\widetilde{\mathbf{U}}=(\widetilde{\boldsymbol{u}}_1,\ldots,\widetilde{\boldsymbol{u}}_M)\in\mathbb{R}^{p\times M}$ denote the centered, RMS-standardized basis, and let $\lambda_1,\ldots,\lambda_M$ be the eigenvalues associated with its untransformed source vectors. All spectral generators use $\gamma=0.9$ and weights $w_j=(1-\gamma\lambda_j)^{-1/2}$. Each replication first generates a latent field $\boldsymbol{s}=(s_1,\ldots,s_p)^{\top}$ and subsequently generates counts conditional on that field.

The mechanisms below standardize each realized field to a prescribed empirical standard deviation $\sigma_s$. This controls realized signal amplitude across scenarios, rather than merely controlling its prior expectation. Where winsorization is applied, values below and above the empirical 0.5th and 99.5th percentiles are replaced by the corresponding percentile values before centering and rescaling. These transformations distinguish the generating distributions from the Gaussian priors used for fitting.

\subsubsection{Sparse RENeGe truth}
\label{sec:truth-sparse}

The sparse mechanism generates a field in the retained spectral span with exactly $k$ active coefficients. In each replication, an active set $\mathcal{A}_0$ of size $k$ is randomly selected from the $M$ candidates. Initial coefficients are $\beta_j^{\ast}=\varepsilon_j A_jw_j$ for $j\in\mathcal{A}_0$ and zero otherwise, where the signs $\varepsilon_j\in\{-1,+1\}$ have equal probabilities and $A_j\sim\textsf{Uniform}(0.7,1.3)$. The raw field $\boldsymbol{s}^{\ast}=\widetilde{\mathbf{U}}\boldsymbol{\beta}^{\ast}$ is centered and rescaled to have empirical standard deviation $\sigma_s$. This rescaling preserves its generating active set.

The easy, moderate, and hard conditions use $k=3$, $6$, and $10$, respectively. This mechanism supplies known field values and component labels for evaluating recovery. It favors the proposed model's representational structure but does not draw coefficients from its fitted continuous-mixture prior: inactive coefficients are exactly zero, and active magnitudes are generated and jointly rescaled as described above.

\subsubsection{Dense CAR-type spectral truth}
\label{sec:truth-car}

The dense mechanism initially generates $\boldsymbol{s}^{\ast}=\widetilde{\mathbf{U}}\boldsymbol{\beta}^{\ast}$ with coefficients $\beta_j^{\ast}=w_j\varepsilon_j$, where $\varepsilon_j\overset{\textsf{iid}}{\sim}\textsf{N}(0,1)$ for $j=1,\ldots,M$. All candidate modes therefore contribute almost surely before transformation. The field is then winsorized, centered, and rescaled to empirical standard deviation $\sigma_s$.

This scenario assesses recovery under dense spectral variation. However, winsorization is nonlinear and can introduce variation outside the retained span, while realization-specific rescaling changes the Gaussian generating distribution. The scenario is therefore CAR-type rather than an exact draw from the fitted spectral CAR model, and sparse active-set recovery is not evaluated.

\subsubsection{BYM2 truth}
\label{sec:truth-bym2}

The BYM2-type mechanism combines structured and independent segment-level variation. A structured field is first generated as
\begin{equation}
  \boldsymbol{s}^{(S)}
  =
  \frac{
    \widetilde{\mathbf{U}}
    (w_1\varepsilon_1,\ldots,w_M\varepsilon_M)^{\top}
  }{
    \big(\sum_{j=1}^{M}w_j^2\big)^{1/2}
  },
  \qquad
  \varepsilon_j\overset{\textsf{iid}}{\sim}\textsf{N}(0,1),
  \label{eq:simulation-structured-field}
\end{equation}
and standardized to unit empirical standard deviation. Independently, $s_e^{(U)}\overset{\textsf{iid}}{\sim}\textsf{N}(0,1)$ is generated for each segment. The raw combined field is $s_e^{\ast}=\sqrt{\rho_0}\,s_e^{(S)}+\sqrt{1-\rho_0}\,s_e^{(U)}$, with $\rho_0=0.20$. It is subsequently winsorized, centered, and rescaled to standard deviation $\sigma_s$.

The substantial independent component challenges models restricted to the low-rank spectral span. Because the structured realization is standardized and the combined field is transformed, this mechanism is not an exact draw from the fitted BYM2 model. In particular, $\rho_0$ specifies the pre-transformation mixture weights and should not be interpreted as an exact structured variance fraction in the final realized field.

\subsubsection{Localized hotspot truth}
\label{sec:truth-hotspot}

The localized mechanism generates nonlinear patterns in the two-dimensional embedding $\boldsymbol{q}_e=(\widetilde u_{e1},\widetilde u_{e2})^{\top}$. Two segments, $c_1$ and $c_2$, are randomly selected as centers, with squared distances $d_{re}=\|\boldsymbol{q}_e-\boldsymbol{q}_{c_r}\|^2$ for $r=1,2$. The raw field is
\begin{equation}
  s_e^{\ast}
  =
  \textsf{exp}\!\left(-\frac{d_{1e}}{2h}\right)
  -
  0.7\,\textsf{exp}\!\left(-\frac{d_{2e}}{2h}\right),
  \label{eq:simulation-hotspots}
\end{equation}
where $h$ is a squared-distance scale determined adaptively from the empirical distances, with a lower bound of 0.05. The field is then winsorized, centered, and rescaled to standard deviation $\sigma_s$. The exact adaptive rule for $h$ must be specified to reproduce this generator.

This construction produces positive and negative localized deviations in spectral coordinates, which need not correspond to geographically compact regions or admit a sparse linear representation in the retained basis. It provides a limited misspecification test because its geometry still derives from that basis. Full-rank CAR and BYM2 generators, and hotspots defined independently through geographic locations or road corridors, were not considered.

\subsection{Exposure}
\label{sec:simulation-exposure}

Let $L_e^{(0)}$ denote the observed length of segment $e$ in kilometres. Lengths are winsorized at their empirical 1st and 99th percentiles, yielding $\widetilde L_e$, with mean $\overline L=p^{-1}\sum_{e=1}^{p}\widetilde L_e$. Simulated exposure is defined by
\begin{equation}
  L_e
  =
  \overline L\,
  \frac{
    (\widetilde L_e/\overline L)^{a_L}
  }{
    p^{-1}\sum_{r=1}^{p}(\widetilde L_r/\overline L)^{a_L}
  }.
  \label{eq:simulation-exposure}
\end{equation}
This preserves the mean winsorized exposure within each network while changing its heterogeneity. Setting $a_L=0$ gives homogeneous exposure $L_e=\overline L$, $a_L=1$ recovers the winsorized lengths, and $a_L=2$ amplifies their relative differences. These are synthetic exposure profiles based on the observed lengths; their total need not equal the original total before winsorization.

\subsection{Crash intensity and count generation}
\label{sec:simulation-counts}

For each retained network, the baseline log frequency is calibrated as $\alpha_{\textsf{base}}=\textsf{log}\{(\sum_{e=1}^{p}Y_e^{\textsf{obs}}+0.5)/\sum_{e=1}^{p}L_e^{(0)}\}$. A scenario-specific multiplier $c$ gives $\alpha_{\textsf{true}}=\alpha_{\textsf{base}}+\textsf{log}(c)$ and expected counts $\mu_e=L_e\,\textsf{exp}(\alpha_{\textsf{true}}+s_e)$. The multiplier changes the baseline frequency conditional on the generated field; it does not fix the realized network-wide mean frequency, which also depends on exposure and $\textsf{exp}(s_e)$.

For each latent field, training and test counts are generated independently conditional on the common means and dispersion:
\begin{equation}
  Y_e^{\textsf{train}},\,Y_e^{\textsf{test}}
  \mid\mu_e,\phi
  \overset{\textsf{ind}}{\sim}
  \textsf{NB}_2(\mu_e,\phi),
  \qquad e=1,\ldots,p.
  \label{eq:simulation-count-generation}
\end{equation}
Counts are also conditionally independent across segments. The parameterization satisfies $\textsf{E}(Y_e\mid\mu_e,\phi)=\mu_e$ and $\textsf{Var}(Y_e\mid\mu_e,\phi)=\mu_e+\mu_e^2/\phi$. Only $\boldsymbol{Y}^{\textsf{train}}$ is used for fitting; $\boldsymbol{Y}^{\textsf{test}}$ is reserved for predictive evaluation. The target is a new count realization on the same segments under the same latent field, rather than prediction on spatially withheld segments.

\subsection{Global difficulty levels}
\label{sec:simulation-difficulty}

The three difficulty levels jointly vary baseline intensity, exposure heterogeneity, signal amplitude, and negative-binomial dispersion (Table~\ref{tab:simulation-difficulty}). Under sparse truth, they also vary the number of active modes. Increasing difficulty reduces intensity and spatial signal amplitude, increases exposure heterogeneity and conditional overdispersion, and makes the sparse representation less parsimonious. These composite conditions are stress tests rather than a factorial design for isolating individual factor effects.

\begin{table}[htbp]
  \centering
  \caption{Composite simulation conditions. The active-set size $k$ applies only to sparse RENeGe truth.}
  \label{tab:simulation-difficulty}
  \begin{tabular}{lccccc}
    \toprule
    Difficulty & $c$ & $a_L$ & $\sigma_s$ & $\phi$ & $k$ \\
    \midrule
    Easy     & 1.50 & 0 & 0.70 & 2.0 & 3 \\
    Moderate & 1.00 & 1 & 0.50 & 1.0 & 6 \\
    Hard     & 0.67 & 2 & 0.35 & 0.5 & 10 \\
    \bottomrule
  \end{tabular}
\end{table}

\subsection{Competing models}
\label{sec:simulation-competitors}

Each training dataset is analyzed using Sparse RENeGe, spectral CAR, spectral BYM2, and the nonspatial negative-binomial baseline. The spatial models receive identical counts, exposures, candidate columns, and source eigenvalues; the nonspatial model uses the same counts and exposures but excludes latent spatial effects. The Sparse RENeGe spike standard deviation is fixed at $\sigma_0=0.05$. Comparisons therefore concern the fitted regularization and heterogeneity structures within the common design.

\subsection{Posterior computation}
\label{sec:simulation-computation}

Models are fitted in Stan through \texttt{cmdstanr}, using two chains with 750 warm-up and 750 post-warm-up iterations per chain. The sampler settings are \texttt{adapt\_delta = 0.95} and \texttt{max\_treedepth = 12}. Computationally intensive performance calculations use at most 300 approximately equally spaced posterior draws from the combined post-warm-up sample. Every simulation uses the complete connected network retained by preprocessing, with no additional segment subsampling. These shorter simulation runs differ from the four-chain application fits described in Section~\ref{sec:computation}.

\subsection{Recovery of the latent spatial field}
\label{sec:field-recovery}

For each spatial model, let $\widehat s_e=\textsf{E}\big(s_e\mid\boldsymbol{Y}^{\textsf{train}}\big)$ denote the posterior mean latent effect. For BYM2, the evaluated effect includes both structured and independent components, matching the total field used to generate counts. Recovery is measured by
\begin{equation}
  \textsf{RMSE}_s
  =
  \left\{
    \frac{1}{p}\sum_{e=1}^{p}(\widehat s_e-s_e)^2
  \right\}^{1/2}.
  \label{eq:simulation-rmse}
\end{equation}
The nonspatial model has $\widehat s_e=0$ and provides a reference for the magnitude of the generating field. Because $\sigma_s$ decreases across difficulty levels, absolute RMSE should be interpreted relative to this baseline: a smaller error in a harder condition need not imply more informative recovery.

Uncertainty is assessed by the segment-averaged coverage $\textsf{Coverage}_{0.90}=p^{-1}\sum_{e=1}^{p}\textsf{I}\{s_e\in[q_{0.05,e},q_{0.95,e}]\}$, where $q_{0.05,e}$ and $q_{0.95,e}$ are marginal posterior quantiles. Averaging this quantity across replications evaluates coverage under the specified generator. It does not assess simultaneous coverage of the complete network field.

\subsection{Recovery of active eigenvectors}
\label{sec:eigenvector-recovery}

Component recovery is evaluated only under sparse RENeGe truth, where the generating active set $\mathcal{A}_0$ is known. Let $\omega_j=\textsf{P}\big(z_j=1\mid\boldsymbol{Y}^{\textsf{train}}\big)$ denote posterior slab membership, and define $\widehat{\mathcal{A}}=\{j:\omega_j\geq0.5\}$. Ranking performance is measured by the area under the receiver operating characteristic curve (AUC), while threshold performance is measured by
\begin{equation}
  \textsf{TPR}
  =
  \frac{|\widehat{\mathcal{A}}\cap\mathcal{A}_0|}{|\mathcal{A}_0|},
  \qquad
  \textsf{FDP}
  =
  \frac{|\widehat{\mathcal{A}}\setminus\mathcal{A}_0|}
       {\max\{|\widehat{\mathcal{A}}|,1\}}.
  \label{eq:simulation-selection-metrics}
\end{equation}
The false discovery proportion (FDP) is defined as zero when no modes are selected, and its mean over replications estimates the false discovery rate (FDR). The handling of empty selections in the implementation should be checked against this convention. We also compare $\textsf{E}\big(K\mid\boldsymbol{Y}^{\textsf{train}}\big)=\sum_{j=1}^{M}\omega_j$ with the generating size $k$.

These metrics evaluate slab membership as a classifier of nonzero generating coefficients. They do not imply that the fitted continuous-mixture model has exact zeros or that its expected membership count must equal the generating active-set size. Their interpretation also remains subject to the spike--slab scale-ordering qualification in Section~\ref{sec:spike-slab}.

\subsection{Out-of-sample predictive performance}
\label{sec:simulation-prediction}

Prediction is evaluated on the independent test realization. For posterior draw $b$, let $\mu_e^{(b)}=L_e\,\textsf{exp}\big(\alpha_0^{(b)}+s_e^{(b)}\big)$. The marginal posterior predictive mass at the observed test count is approximated by
\begin{equation}
  \widehat p\big(
    Y_e^{\textsf{test}}\mid\boldsymbol{Y}^{\textsf{train}}
  \big)
  =
  \frac{1}{B}\sum_{b=1}^{B}
  f_{\textsf{NB}_2}\big(
    Y_e^{\textsf{test}};\mu_e^{(b)},\phi^{(b)}
  \big),
  \label{eq:simulation-predictive-density}
\end{equation}
where $f_{\textsf{NB}_2}$ denotes the negative-binomial probability mass function. For BYM2, $s_e^{(b)}$ includes the segment-specific effect learned from the training count. This is appropriate because the test realization shares that segment's latent field; it differs from the leave-one-segment-out target used in the applications.

The realized test log score is $\widehat{\textsf{ELPD}}_{\textsf{test}}=\sum_{e=1}^{p}\textsf{log}\,\widehat p\big(Y_e^{\textsf{test}}\mid\boldsymbol{Y}^{\textsf{train}}\big)$. Its per-segment version, $\overline{\textsf{ELPD}}_{\textsf{test}}=p^{-1}\widehat{\textsf{ELPD}}_{\textsf{test}}$, removes the direct dependence on network size, although absolute scores still depend on the generating count distribution. Averaging across replications estimates expected predictive performance under each scenario. This sum of marginal log scores is not the joint log predictive density of the entire test vector.

For graphical comparison within each dataset, model $m$ is assigned $\Delta_m=\overline{\textsf{ELPD}}_{\textsf{test},m}-\max_{m'}\overline{\textsf{ELPD}}_{\textsf{test},m'}$. The best realized score is therefore zero and all others are nonpositive. These relative scores describe performance against the dataset-specific winner, rather than against a fixed reference model.

\subsection{Computational diagnostics}
\label{sec:simulation-diagnostics}

Computational summaries include elapsed time, numbers of divergent fits and transitions, minimum bulk effective sample size, and the proportion of fits satisfying a joint diagnostic rule. A fit passes when $N_{\textsf{div}}=0$, $N_{\textsf{treedepth}}=0$, and $\min(\textsf{ESS}_{\textsf{bulk}})\geq200$, where the first two quantities count divergent transitions and transitions reaching the maximum tree depth. This rule contains no $\widehat R$ cutoff and should not be interpreted as a comprehensive convergence criterion. The monitored parameter set also affects the minimum ESS and must be held explicit when interpreting comparisons.

Execution failures are distinguished from completed fits with unfavorable diagnostics. All completed fits are retained in the performance summaries, so diagnostic reporting identifies computational limitations but does not eliminate their possible influence on estimated model performance.

\subsection{Monte Carlo replication and reproducibility}
\label{sec:simulation-reproducibility}

Each of the 24 scenarios has 50 independent replications. Simulation tasks use the deterministic seed rule $\textsf{seed}=910000+1000\times\textsf{scenario\_id}+\textsf{replicate}$, with model-specific sampling seeds derived from each task seed. Distinct seeds support reproducibility but do not by themselves establish formally independent random-number streams; exact reproduction additionally requires the seed-derivation rule and generator implementation.

Within each scenario and fitted model, performance is summarized by its empirical mean and standard deviation across replications. The Monte Carlo standard error of a mean is $\textsf{MCSE}=s_R/\sqrt{R}$, where $R=50$ and $s_R$ is the empirical standard deviation. Paired model comparisons use within-replication differences because all models are fitted to the same generated data. Pooled comparisons additionally combine scenarios; their reported standard errors are retained from the available summaries and cannot be reconstructed from the aggregate tables alone. The workflow proceeds from a fixed network basis to a generated latent field, independent training and test counts, model fitting, and evaluation of recovery, uncertainty, prediction, and computation.

\subsection{Simulation results}
\label{sec:simulation-results}

All 1,200 datasets were generated, and every model returned a fit for every dataset, yielding 4,800 completed executions without fitting errors. Successful execution did not imply satisfactory posterior computation. The statistical summaries below include all 50 replications in every network--mechanism--difficulty cell and must therefore be interpreted alongside the diagnostic results.

Table~\ref{tab:simulation-performance} pools the two networks and three difficulty levels within each generating mechanism, giving 300 datasets per mechanism--model combination. Under sparse RENeGe truth, the proposed model had the lowest mean field RMSE (0.0996) and highest mean test log score ($-2.2734$). Relative to spectral CAR, its mean paired RMSE improvement was 0.0020 (MCSE 0.00065), and its mean paired log-score improvement was 0.00018 (MCSE 0.00003); these differences use unrounded summaries. Spectral CAR performed best under the dense CAR-type generator, while spectral BYM2 performed best under the BYM2-type and localized generators. These rankings broadly followed the generating structure, although the transformations described above prevent treating the dense generators as exactly specified fitted models.

\begin{table}[htbp]
  \centering
  \small
  \caption{Performance averaged over both networks and all three difficulty levels. Each mechanism--model entry summarizes 300 datasets. Lower field RMSE and higher test mean log score indicate better performance; bold values identify the best reported mean within each mechanism.}
  \label{tab:simulation-performance}
  \begin{tabular}{@{}llrr@{}}
    \toprule
    Generating mechanism & Fitted model & Field RMSE & Test mean log score \\
    \midrule
    BYM2-type & Sparse RENeGe & 0.4731 & $-2.3742$ \\
              & Spectral CAR & 0.4731 & $-2.3741$ \\
              & Spectral BYM2 & \textbf{0.4297} & $\mathbf{-2.3480}$ \\
              & Nonspatial NB & 0.5166 & $-2.3927$ \\
    \addlinespace
    Dense CAR-type & Sparse RENeGe & 0.1152 & $-2.2767$ \\
                   & Spectral CAR & \textbf{0.1119} & $\mathbf{-2.2766}$ \\
                   & Spectral BYM2 & 0.1155 & $-2.2769$ \\
                   & Nonspatial NB & 0.5166 & $-2.3983$ \\
    \addlinespace
    Localized hotspot & Sparse RENeGe & 0.2464 & $-2.3025$ \\
                      & Spectral CAR & 0.2491 & $-2.3026$ \\
                      & Spectral BYM2 & \textbf{0.2427} & $\mathbf{-2.2991}$ \\
                      & Nonspatial NB & 0.5166 & $-2.4065$ \\
    \addlinespace
    Sparse RENeGe & Sparse RENeGe & \textbf{0.0996} & $\mathbf{-2.2734}$ \\
                  & Spectral CAR & 0.1015 & $-2.2736$ \\
                  & Spectral BYM2 & 0.1056 & $-2.2740$ \\
                  & Nonspatial NB & 0.5166 & $-2.4176$ \\
    \bottomrule
  \end{tabular}
\end{table}

\begin{figure}[htbp]
  \centering
  \includegraphics[width=\textwidth]{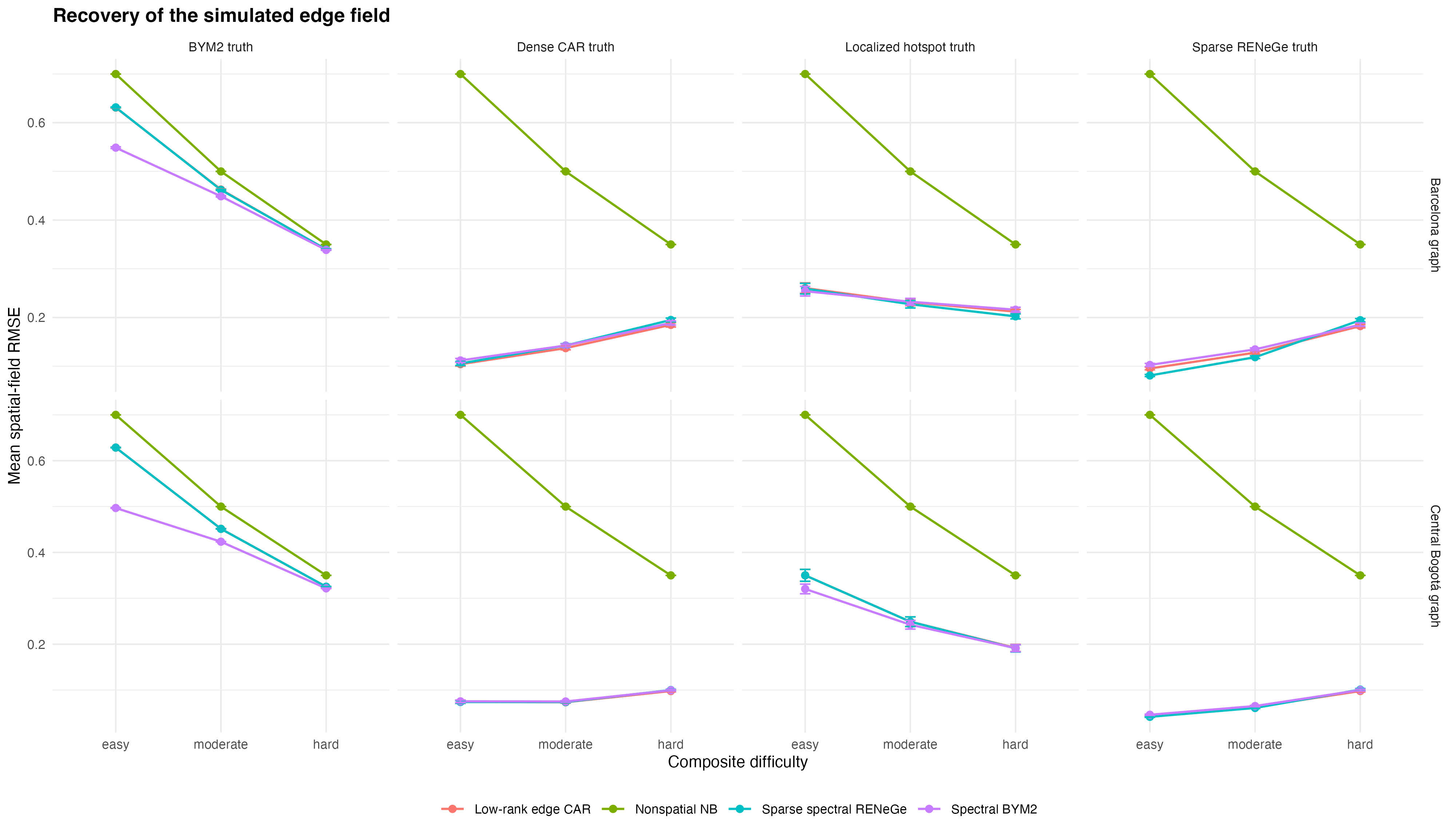}
  \caption{Mean latent-field RMSE across 50 replications for each network, generating mechanism, and composite difficulty level. Error bars denote Monte Carlo standard errors. Signal amplitude decreases across difficulty levels, so absolute errors should be interpreted relative to the nonspatial baseline.}
  \label{fig:simulation-rmse}
\end{figure}

The difficulty trajectories in Fig.~\ref{fig:simulation-rmse} show similar field recovery for Sparse RENeGe and spectral CAR in many cells, consistent with their common candidate basis. The nonspatial baseline reflects the imposed signal amplitude rather than fitted spatial recovery. For a centered field standardized using sample variance with denominator $p-1$, its zero-field RMSE is $\sigma_s\sqrt{(p-1)/p}$. Averaging this quantity over the three amplitudes and the two network sizes is consistent with the reported value 0.5166.

Average coverage of nominal 90\% marginal credible intervals for Sparse RENeGe was 88.7\% under sparse truth and 85.8\% under dense CAR-type truth, falling to 54.4\% under localized truth and 25.0\% under BYM2-type truth. Coverage was therefore closest to nominal under sparse, basis-compatible signals but deteriorated substantially when the field contained variation poorly represented by the retained basis. Comparable predictive scores should not be taken as evidence of reliable latent-field uncertainty under these forms of misspecification.

\begin{figure}[htbp]
  \centering
  \includegraphics[width=\textwidth]{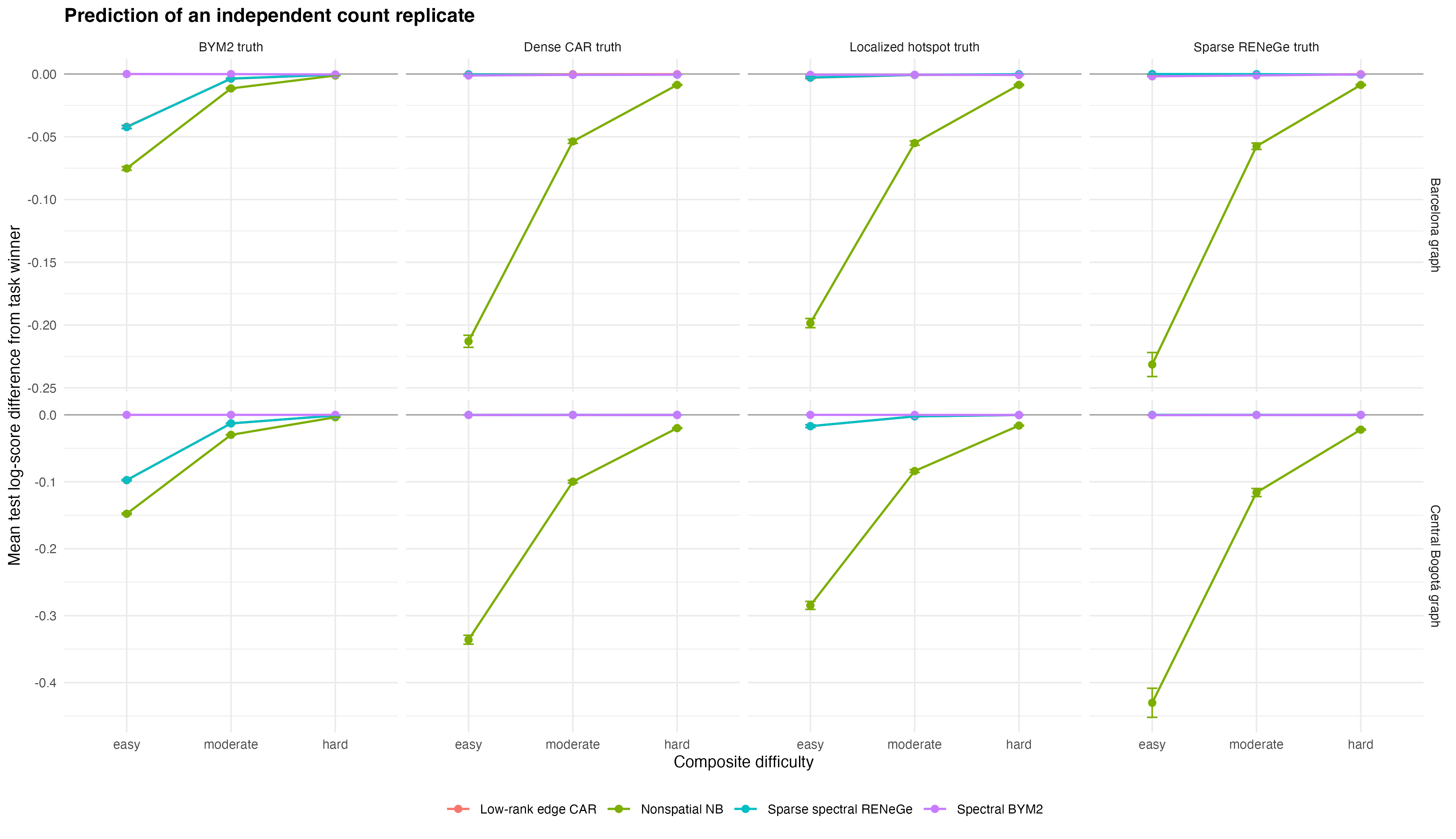}
  \caption{Mean per-segment test log-score difference from the best-scoring model within each dataset, evaluated on an independent count realization sharing the training latent field. Zero denotes the dataset-specific winner; error bars denote Monte Carlo standard errors.}
  \label{fig:simulation-prediction}
\end{figure}

Predictive differences among spatial models were generally smaller than differences in field recovery (Fig.~\ref{fig:simulation-prediction}). Averaged over the 300 datasets for each mechanism, Sparse RENeGe improved the test mean log score over the nonspatial baseline by 0.1442 under sparse truth, 0.1215 under dense CAR-type truth, 0.1040 under localized truth, and 0.0186 under BYM2-type truth. The corresponding reported paired MCSEs were 0.0095, 0.0069, 0.0057, and 0.0010. Thus, a restricted spectral representation retained predictive value under the considered misspecifications, without uniformly outperforming the more closely aligned comparator.

\begin{figure}[htbp]
  \centering
  \includegraphics[width=0.88\textwidth]{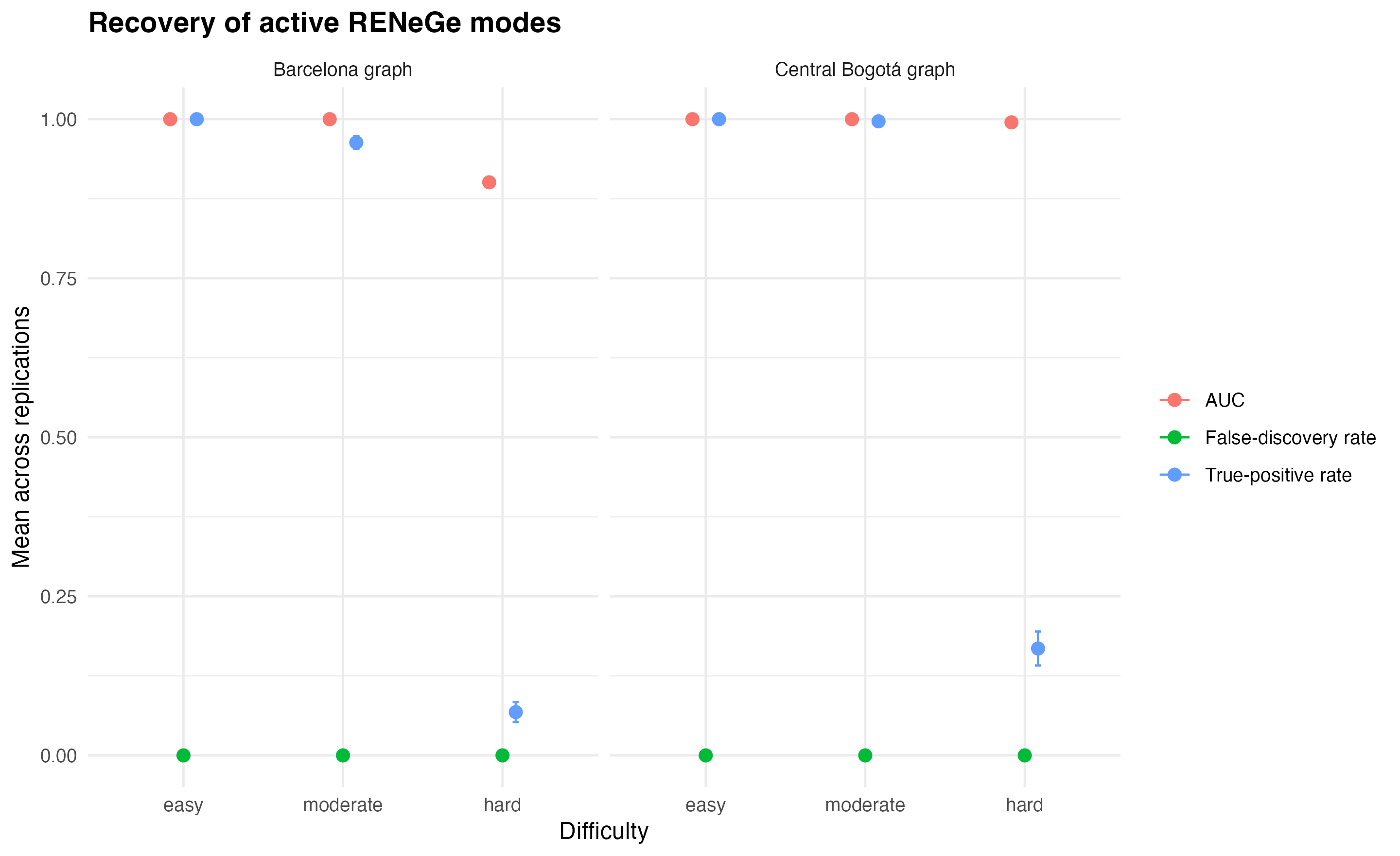}
  \caption{Mean AUC, true-positive rate, and false discovery proportion under sparse RENeGe truth. Error bars denote Monte Carlo standard errors across 50 replications. Threshold-based metrics use posterior slab-membership probability at least 0.5.}
  \label{fig:simulation-selection}
\end{figure}

Under sparse truth, ranking and threshold-based recovery differed substantially across difficulty levels (Table~\ref{tab:simulation-selection} and Fig.~\ref{fig:simulation-selection}). Mean AUC rounded to 1.000 in both networks under easy and moderate conditions, with moderate-condition TPRs of 0.963 in Barcelona and 0.997 in central Bogot\'a. The reported mean FDR was 0.000 in every cell. Under hard conditions, AUC remained 0.901 and 0.995, respectively, but TPR fell to 0.068 and 0.168. These results indicate that active components could remain highly ranked while few exceeded the fixed 0.5 membership threshold.

Posterior expected membership counts also differed from threshold-selected counts. In moderate conditions, the mean expected counts exceeded the generating value of six, despite the reported near-zero FDR, because membership probabilities below 0.5 still contribute to $\textsf{E}(K\mid\boldsymbol{Y}^{\textsf{train}})$. Under hard conditions, expected counts of 2.71 and 4.59 were well below the generating value of ten. Membership probabilities, rankings, and threshold decisions therefore convey complementary information and should not be treated as interchangeable measures of recovery.

\begin{table}[htbp]
  \centering
  \small
  \caption{Component recovery by Sparse RENeGe under sparse truth, averaged over 50 replications. Here, $k$ is the generating active-set size, FDR is the mean false discovery proportion, and the final column averages posterior expected slab-membership counts.}
  \label{tab:simulation-selection}
  \begin{tabular}{@{}llrrrrr@{}}
    \toprule
    Network & Difficulty & $k$ & AUC & TPR & FDR &
    $\textsf{E}(K\mid\boldsymbol{Y}^{\textsf{train}})$ \\
    \midrule
    Barcelona & Easy & 3 & 1.000 & 1.000 & 0.000 & 3.48 \\
              & Moderate & 6 & 1.000 & 0.963 & 0.000 & 7.54 \\
              & Hard & 10 & 0.901 & 0.068 & 0.000 & 2.71 \\
    Central Bogot\'a & Easy & 3 & 1.000 & 1.000 & 0.000 & 3.42 \\
                     & Moderate & 6 & 1.000 & 0.997 & 0.000 & 7.58 \\
                     & Hard & 10 & 0.995 & 0.168 & 0.000 & 4.59 \\
    \bottomrule
  \end{tabular}
\end{table}

Computational diagnostics limit the strength of the comparisons (Table~\ref{tab:simulation-diagnostics}). The reported joint pass rates were 90.3\% for Sparse RENeGe, 67.0\% for spectral CAR, and 10.0\% for spectral BYM2. Sparse RENeGe produced 105 divergent transitions across 69 fits; BYM2 produced 498 across 75 fits, with a maximum of 47 in one fit. Spectral CAR had no divergences, so failures under the stated rule must involve tree-depth saturation or insufficient minimum bulk ESS. The available aggregates do not allow the pass classifications or their individual causes to be reconstructed.

The mean minimum bulk ESS was 1,295 for Sparse RENeGe, 336 for spectral CAR, and 124 for spectral BYM2. These differences indicate model-dependent computational performance under the common sampling budget; they do not establish that this budget was adequate for every model. In particular, BYM2 comparisons require caution because its low pass rate and limited effective sample sizes can affect both posterior summaries and predictive scores. Model-specific tuning and longer runs are needed to determine the stability of the reported differences.

\begin{table}[htbp]
  \centering
  \small
  \caption{Computational summaries across 1,200 fits per model. Pass rate refers to the joint rule in Section~\ref{sec:simulation-diagnostics}. Divergent fits contain at least one divergent transition; ESS denotes bulk effective sample size, and time is mean elapsed time per fit.}
  \label{tab:simulation-diagnostics}
  \begin{tabular}{@{}lrrrrr@{}}
    \toprule
    Model & Pass (\%) & Divergent fits & Total divergences &
    Mean min.\ ESS & Time (s) \\
    \midrule
    Sparse RENeGe & 90.3 & 69 & 105 & 1,295 & 16.6 \\
    Spectral CAR & 67.0 & 0 & 0 & 336 & 24.5 \\
    Spectral BYM2 & 10.0 & 75 & 498 & 124 & 87.7 \\
    Nonspatial NB & 98.7 & 0 & 0 & 965 & 5.5 \\
    \bottomrule
  \end{tabular}
\end{table}

Within the considered design, Sparse RENeGe combined accurate field recovery and informative component rankings under sparse truth with predictive performance close to spectral CAR. Additional independent effects improved BYM2 performance when the generating field contained substantial segment-specific variation, although its computational diagnostics were substantially weaker. The shared candidate basis, transformed generators, composite difficulty conditions, and retained diagnostically unfavorable fits constrain broader claims about robustness or model superiority.

\section{Barcelona Results}
\label{sec:barcelona-results}

The marginalized spike-and-slab model was fitted to the 2,540 segments in Barcelona's largest connected component. All four chains completed without divergent transitions, and the maximum observed tree depth was 5, below the specified limit of 13. Across the intercept, dispersion, slab-scale, and inclusion-rate parameters, the maximum $\widehat R$ was 1.0018 and the minimum bulk effective sample size was 1,456. These summaries support satisfactory sampling of the monitored parameters. Posterior means were $\textsf{E}(\phi\mid\boldsymbol{y})=0.488$ and $\textsf{E}(\tau\mid\boldsymbol{y})=0.0747$. The dispersion estimate indicates appreciable conditional extra-Poisson variation under the fitted model, whose magnitude also depends on the segment-specific mean.

Modes 3 and 16 had the highest posterior slab-membership probabilities, approximately 0.94 each, followed by mode 2 at approximately 0.80. Modes 9 and 14 had probabilities near 0.50 and 0.48, respectively, making threshold-based classification sensitive to small differences in the estimated probabilities. Values are reported approximately because the available summaries differ in their final digits. The posterior expected membership count was $\textsf{E}(K\mid\boldsymbol{y})=6.80$. This summarizes allocation to the slab across the 20 candidates; it does not identify a model with exactly seven nonzero coefficients or describe the full posterior uncertainty in $K$.

\begin{figure}[htbp]
  \centering
  \includegraphics[width=0.85\textwidth]{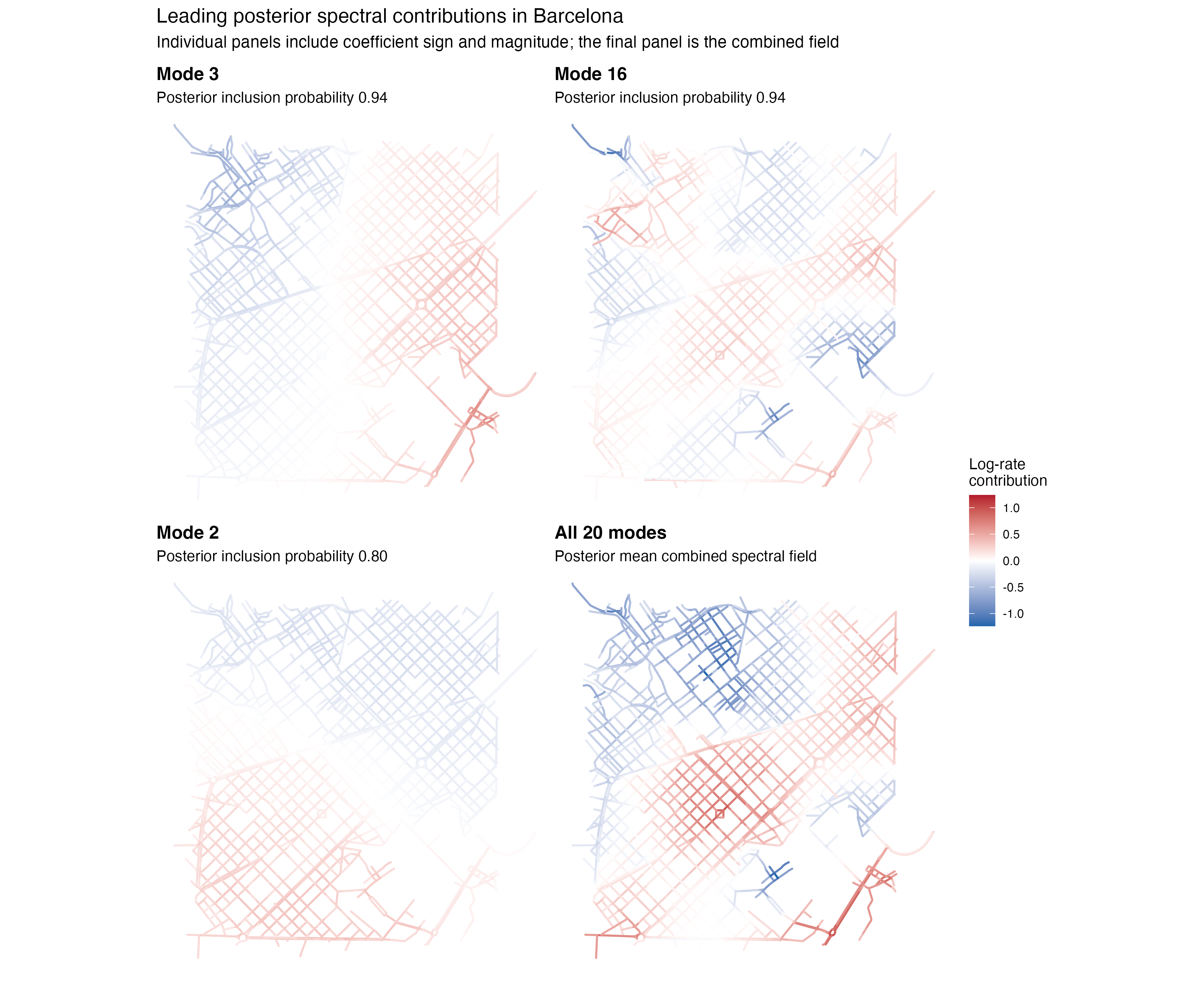}
  \caption{Barcelona posterior mean log-frequency contributions from the three candidates with the highest slab-membership probabilities and the combined field from all 20 candidates. The panels summarize additive spatial effects, not posterior mean crash frequencies.}
  \label{fig:barcelona-contributions}
\end{figure}

Figure~\ref{fig:barcelona-contributions} displays the posterior mean contributions of the three candidates with the highest slab-membership probabilities alongside the combined field from all 20 candidates. Mode 3 primarily contrasts the northwest with the eastern and southeastern network, mode 16 contributes finer alternating contrasts, and mode 2 describes a broad southwest--northeast division. These are signed contributions to log frequency relative to the model intercept. Their interpretation depends on the coefficient-weighted basis functions, whose contributions are invariant to arbitrary eigenvector sign reversals. Ranking by slab membership does not necessarily rank contribution magnitude or identify individual segments with elevated crash frequency.

\begin{figure}[htbp]
  \centering
  \includegraphics[
    width=0.72\textwidth,
    height=0.65\textheight,
    keepaspectratio
  ]{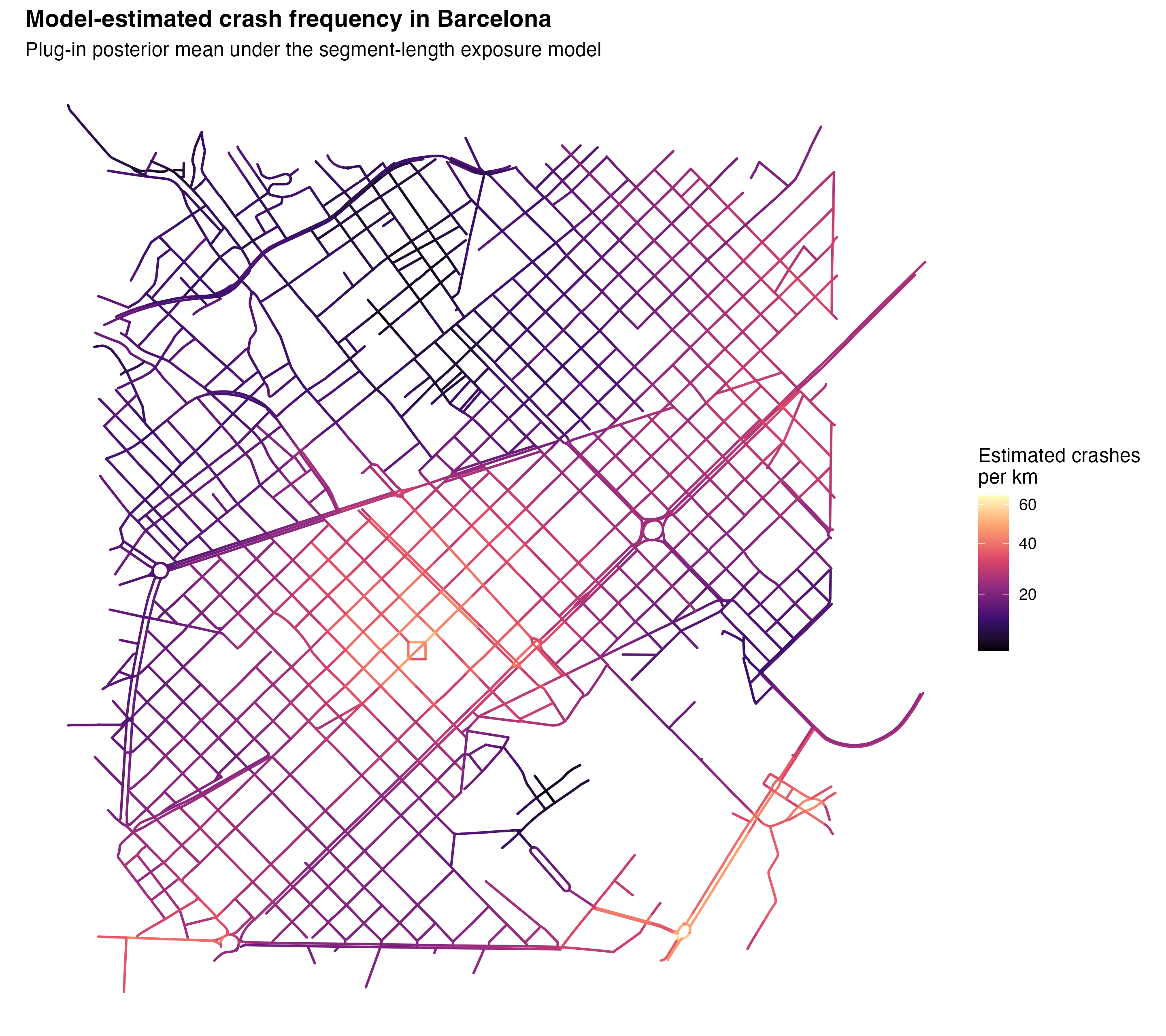}
  \caption{Barcelona fitted crash frequency per kilometre over the observation period, evaluated at posterior mean parameters. The surface includes all 20 candidate contributions and represents a plug-in summary rather than a posterior mean or traffic-adjusted risk estimate.}
  \label{fig:barcelona-rate}
\end{figure}

Figure~\ref{fig:barcelona-rate} combines the intercept and all candidate contributions to display fitted crash frequency per kilometre over the observation period, evaluated at posterior mean parameters. This plug-in surface describes the fitted pattern on the retained network. It differs from posterior mean frequency because exponentiation is nonlinear, and it displays neither posterior uncertainty nor traffic-volume-adjusted risk.

\section{Bogot\'a Results}
\label{sec:bogota-results}

The same model structure and hyperparameter settings were applied to the 6,076 segments in the largest connected component of central Bogot\'a, with the intercept prior centered using this dataset. The four chains completed without divergent transitions, and the maximum observed tree depth was 5. Across the monitored core parameters, the maximum $\widehat R$ was 1.0013 and the minimum bulk effective sample size was 1,497. Posterior means were $\textsf{E}(\phi\mid\boldsymbol{y})=0.378$ and $\textsf{E}(\tau\mid\boldsymbol{y})=0.0958$, indicating substantial conditional count variation alongside the fitted spectral field.

Mode 7 had the highest posterior slab-membership probability, approximately 0.97, followed by mode 2 at approximately 0.91 and mode 15 at approximately 0.55. The posterior expected membership count was $\textsf{E}(K\mid\boldsymbol{y})=4.44$. As in Barcelona, probabilities are approximate because the available summaries differ slightly. Candidate indices refer to each network's own basis, so equally numbered modes across cities do not represent the same spatial pattern. Differences in expected membership counts likewise do not establish differences in intrinsic spatial complexity, because the bases, observation periods, event definitions, and exposure conditions are not directly comparable.

\begin{figure}[htbp]
  \centering
  \includegraphics[width=0.85\textwidth]{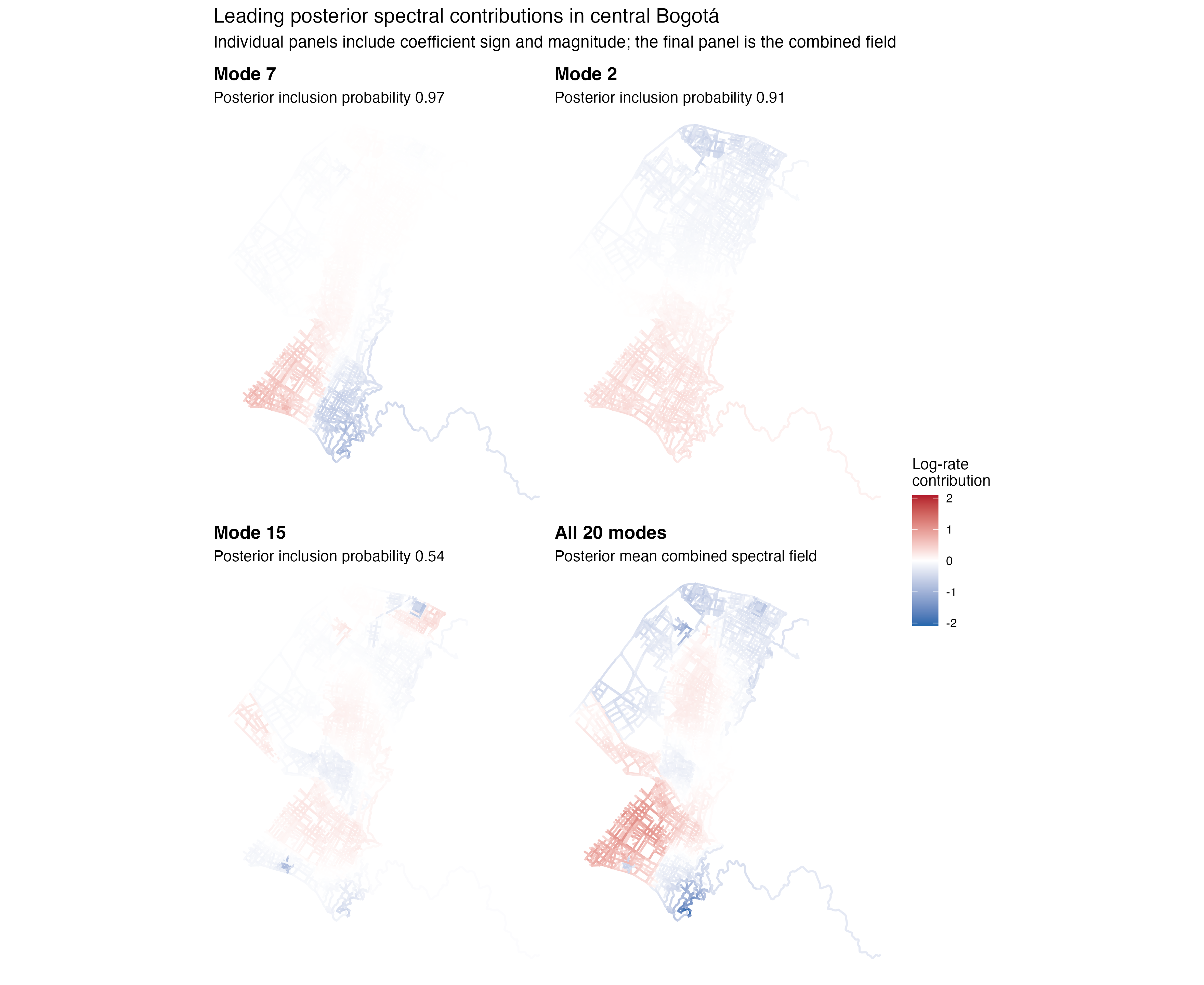}
  \caption{Central Bogot\'a posterior mean log-frequency contributions from the three candidates with the highest slab-membership probabilities and the combined field from all 20 candidates. The panels describe additive spatial effects on the retained component.}
  \label{fig:bogota-contributions}
\end{figure}

The leading contributions in Fig.~\ref{fig:bogota-contributions} describe north--south and central--peripheral contrasts within the retained network. These patterns summarize fitted log-frequency variation and do not identify causal effects of localities or measured road characteristics. The combined panel retains all 20 candidates, including those with low slab-membership probabilities.

\begin{figure}[htbp]
  \centering
  \includegraphics[
    width=0.72\textwidth,
    height=0.65\textheight,
    keepaspectratio
  ]{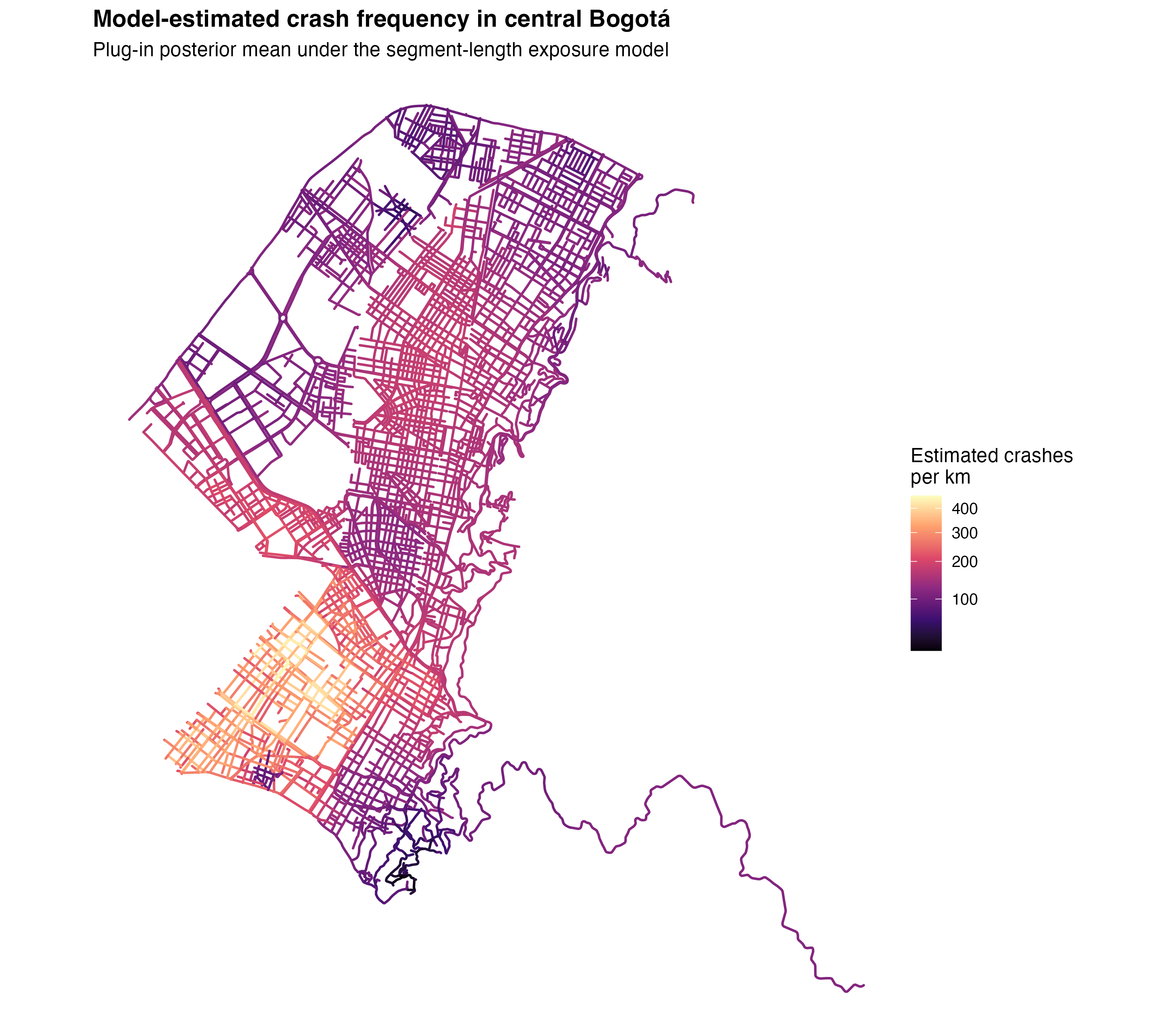}
  \caption{Central Bogot\'a fitted crash frequency per kilometre, evaluated at posterior mean parameters using all 20 candidate contributions. This plug-in surface describes the retained network component and does not quantify posterior uncertainty or traffic-adjusted risk.}
  \label{fig:bogota-rate}
\end{figure}

Figure~\ref{fig:bogota-rate} displays the corresponding plug-in frequency-per-kilometre surface evaluated at posterior mean parameters. Its interpretation is restricted to the modeled component and the observation period represented by the data. As in Barcelona, the map does not display posterior mean frequency, uncertainty, or traffic-adjusted risk; direct comparison of frequency levels between cities requires harmonized observation periods and event definitions.

\section{Predictive Model Comparison}
\label{sec:predictive-comparison}

\begin{figure}[htbp]
  \centering
  \includegraphics[width=0.96\textwidth]{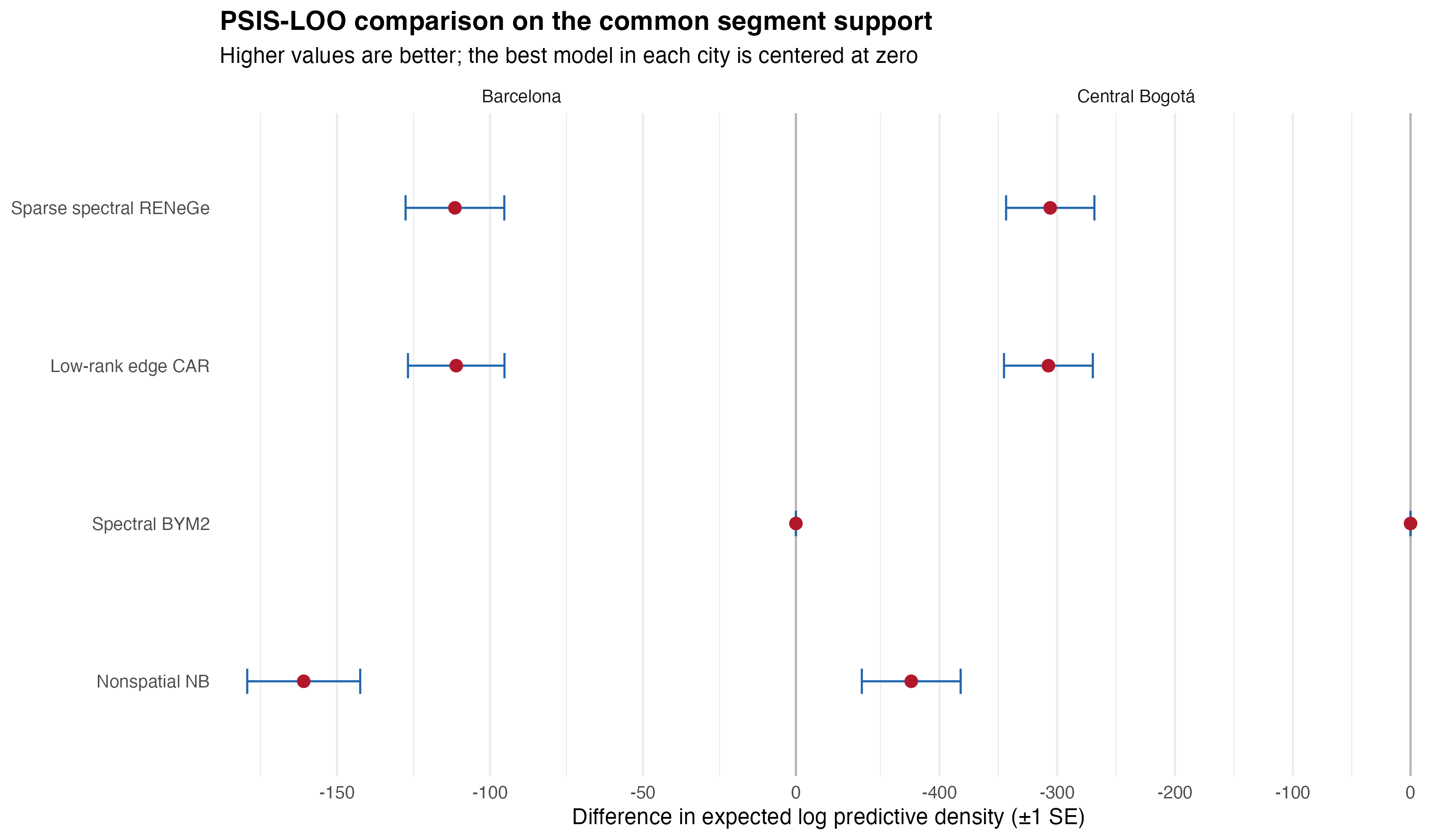}
  \caption{Paired differences in pointwise PSIS-LOO expected log predictive density, with error bars representing one estimated standard error. The BYM2 segment-specific independent effect was integrated against its prior when evaluating the held-out count.}
  \label{fig:loo-comparison}
\end{figure}

Spectral BYM2 achieved the highest exploratory pointwise predictive scores in both applications (Table~\ref{tab:information-criteria} and Fig.~\ref{fig:loo-comparison}). Sparse RENeGe minus BYM2 ELPD differences were $-111.5$ (SE 16.1) in Barcelona and $-305.9$ (SE 37.5) in central Bogot\'a. The BYM2 posterior means of $\rho$ were 0.090 and 0.125, respectively, indicating that the fitted decomposition assigned most average latent variance to the independent component. Its predictive advantage is consistent with heterogeneity not adequately represented by the retained spectral field, although these summaries do not identify its source or separate it fully from negative-binomial overdispersion.

Sparse RENeGe and spectral CAR had similar predictive scores. Their paired ELPD differences, computed as Sparse RENeGe minus CAR, were $-0.42$ (SE 1.76) in Barcelona and $1.55$ (SE 1.58) in central Bogot\'a. These differences provide little evidence of predictive separation at the reported precision, although they do not constitute a formal equivalence assessment. Both models scored above the nonspatial baseline. Within the common candidate basis, the mixture prior therefore supplied component-membership summaries without a clearly resolved gain or loss in pointwise prediction.

\begin{table}[htbp]
  \centering
  \small
  \caption{Exploratory pointwise PSIS-LOO summaries on each retained network component. Higher ELPD and lower LOOIC indicate better scores under the fitted empirical-Bayes prior specification. Metrics are rounded separately from unrounded summaries.}
  \label{tab:information-criteria}
  \resizebox{\textwidth}{!}{%
    \begin{tabular}{llrrrr}
      \toprule
      City & Model & ELPD & SE(ELPD) & $p_{\textsf{LOO}}$ & LOOIC \\
      \midrule
      Barcelona
      & Spectral BYM2 & $\mathbf{-3837.5}$ & 72.5 & 19.1 & $\mathbf{7675.1}$ \\
      & Spectral CAR & $-3948.6$ & 78.7 & 33.9 & $7897.2$ \\
      & Sparse RENeGe & $-3949.0$ & 78.8 & 33.0 & $7898.0$ \\
      & Nonspatial NB & $-3998.4$ & 79.8 & 4.3 & $7996.9$ \\
      \addlinespace
      Central Bogot\'a
      & Spectral BYM2 & $\mathbf{-19835.6}$ & 141.9 & 21.6 & $\mathbf{39671.2}$ \\
      & Sparse RENeGe & $-20141.6$ & 153.7 & 39.3 & $40283.1$ \\
      & Spectral CAR & $-20143.1$ & 153.8 & 41.6 & $40286.2$ \\
      & Nonspatial NB & $-20259.6$ & 154.8 & 4.9 & $40519.2$ \\
      \bottomrule
    \end{tabular}%
  }
\end{table}

All comparator fits completed without divergent transitions. The largest core-parameter $\widehat R$ was 1.0071, and the minimum core bulk effective sample size was 389. No reported Pareto diagnostic exceeded $\widehat k=1$; however, one segment had $\widehat k>0.7$ for Sparse RENeGe in each city, and two did so for the Bogot\'a CAR fit. These observations warrant exact leave-one-out refits or importance-sampling corrections to assess the stability of the affected scores, particularly for the small Sparse RENeGe--CAR differences.

The reported rankings concern withholding one segment count while retaining neighboring observations and network information. The BYM2 independent effect was integrated out for this target, as described in Section~\ref{sec:predictive-assessment}. Full-data centering of the intercept prior, unverified quadrature accuracy, and the absence of spatially blocked validation limit stronger predictive claims. Aggregate ELPD values should also be compared within cities rather than across them, because the networks differ in size and count distributions.

\section{Discussion}
\label{sec:discussion}

The proposed model combines a road-segment spectral representation with continuous-mixture regularization and posterior component-membership summaries. Its empirical contribution is an interpretable decomposition of fitted network variation with predictive scores close to those of spectral CAR. The comparisons do not show a clear predictive benefit from the mixture layer itself. Instead, they illustrate how probabilistic component assessment can complement dense spectral estimation within a fixed candidate space.

The simulation supports this interpretation under the mechanisms considered. Sparse RENeGe had the best average field recovery and test prediction under exactly sparse generating fields, while the dense CAR-type and BYM2-type generators favored their more closely aligned comparators. Under the hardest sparse condition, high AUC coexisted with low threshold sensitivity, demonstrating that ranking and selection at probability 0.5 can yield different conclusions. Likewise, expected slab-membership counts need not match either the generating active-set size or the number of threshold-selected components. These summaries are best reported together, with their dependence on basis choice and mixture-scale ordering made explicit.

Uncertainty calibration was more sensitive to misspecification than point prediction. The proposed model's marginal interval coverage was close to nominal under sparse truth but deteriorated markedly under localized and independently varying fields. This distinction matters for applications: reasonable predictive scores do not guarantee reliable uncertainty about segment-level latent effects. Posterior predictive assessment and sensitivity to basis size, shrinkage specification, and residual heterogeneity are therefore needed before using the fitted surfaces to identify unusually high-frequency segments.

Spectral BYM2 achieved the highest application scores and allocated most average latent variance to its independent component. This suggests that the retained spectral representation leaves substantial local heterogeneity unresolved. Possible explanations include omitted road characteristics, traffic exposure, limitations of the truncated basis, and other segment-specific variation; the fitted decomposition does not distinguish among them. Moreover, the simulation predicts repeated counts sharing an existing latent field, whereas application PSIS-LOO predicts a segment count withheld from fitting. Agreement or disagreement between their rankings must be interpreted in light of these different targets.

Computational limitations also constrain the simulation evidence. Sparse RENeGe passed the stated joint diagnostic rule in 90.3\% of fits, but 69 fits contained divergences. Spectral BYM2 had substantially lower effective sample sizes and a pass rate of only 10.0\%. Because all completed fits entered the performance summaries, model-specific tuning and longer runs are needed to determine the stability of the numerical comparisons. The four-chain application fits had more favorable reported diagnostics, although core-parameter summaries alone do not assess every derived quantity.

The scope of the spatial comparison is deliberately restricted. All structured components use the same truncated, centered, and scaled basis, whose induced covariance differs from an exact RENeGe covariance. Full-rank edge models, fitted nodal alternatives, and other spectral shrinkage priors were not evaluated. Component interpretations are also conditional on the computed basis, particularly when eigenspaces admit multiple equivalent representations. The results consequently support the usefulness of the proposed representation within this design rather than a general advantage of edge-based selection over other spatial approaches.

Substantive interpretation requires additional data and validation. Segment length yields frequency per unit roadway over the observation period, not traffic-adjusted crash risk, and the current fits cannot distinguish exposure-related variation from roadway characteristics. Verified event provenance, documented assignment procedures, harmonized observation periods, and sensitivity to excluded network components are necessary for cross-city comparisons. Data-independent prior assessment, posterior predictive checks, and spatially blocked validation would further clarify model adequacy and predictive scope. Even with these extensions, causal interpretation would require assumptions and a design beyond the present descriptive analysis.

\section{Conclusions}
\label{sec:conclusions}

This study develops a Bayesian count model whose latent spatial field is defined directly on the road segments where crashes are observed. An edge-neighborhood operator supplies candidate spectral patterns, and a continuous spike-and-slab prior provides regularization and posterior membership probabilities. The construction supports interpretable field decompositions while retaining uncertainty about component allocation, without requiring exact-zero coefficients or a single selected model.

In Barcelona and central Bogot\'a, Sparse RENeGe produced differentiated component-membership summaries and pointwise predictive scores close to spectral CAR. Spectral BYM2 scored higher under the reported exploratory leave-one-segment-out assessment, consistent with additional heterogeneity beyond the retained structured field. Simulations favored Sparse RENeGe under sparse generating signals but also revealed sensitivity of interval coverage and threshold-based recovery to misspecification and difficulty. These findings remain conditional on the shared basis, fitted priors, and model-dependent computational diagnostics.

Defining the field on segments aligns latent effects with the observational support, while deterministic projection permits intersection-level displays with propagated posterior uncertainty. Such displays do not establish superiority over fitted nodal models. Further work should verify and harmonize the underlying data, incorporate traffic exposure and measured road attributes, and evaluate prior sensitivity, distributional adequacy, and spatially blocked prediction before translating fitted frequency patterns into intervention priorities.

% ============================================================
% DECLARATIONS
% ============================================================

\section*{Data Availability}
\phantomsection
\addcontentsline{toc}{section}{Data Availability}

Code, processing scripts, derived model inputs, and documentation for constructing the Bogot\'a road-segment network and processing the associated crash records are available at \url{https://github.com/DannaCruz/BogotaAccidentes}. The repository includes example outputs and diagnostic materials from the 2023 analysis. Original source datasets are not redistributed; the documentation identifies the required inputs and describes the procedures for processing, filtering, assigning, and aggregating crash records to road segments. These materials document the Bogot\'a data-processing workflow; they do not, by themselves, establish reproducibility of all model fits, simulations, and results reported in this article.

\section*{Declaration of Competing Interest}
\phantomsection
\addcontentsline{toc}{section}{Declaration of Competing Interest}

The authors declare that they have no known competing financial interests or personal relationships that could have appeared to influence the work reported in this article.

\section*{Declaration of Generative AI and AI-Assisted Technologies in the Manuscript Preparation Process}
\phantomsection
\addcontentsline{toc}{section}{Declaration of Generative AI and AI-Assisted Technologies}

During the preparation of this manuscript, the authors used OpenAI Codex to assist with manuscript organization, code development, and language editing. The authors reviewed and edited the AI-assisted material and take full responsibility for the final content of the article.

% ============================================================
% REFERENCES
% ============================================================

\bibliographystyle{plainnat}
\bibliography{references.bib}

% ============================================================
% APPENDICES
% ============================================================
\clearpage
\appendix

\section{Selected Spectral Bases and Intersection Projections}
\label{app:spectral-bases}

\subsection{Leading spectral contributions}
\label{app:leading-contributions}

Among the fitted models, only Sparse RENeGe provides posterior slab-membership probabilities for individual candidate components. Table~\ref{tab:selected-modes} reports the three candidates with the highest probabilities in each city. These rankings summarize mixture allocation within the retained basis, rather than exact exclusion of other coefficients or ranking by contribution magnitude. Spectral CAR and BYM2 retain all candidate components under Gaussian regularization; their coefficients can be summarized by posterior contribution, but they do not provide analogous mixture-membership probabilities.

\begin{table}[htbp]
  \centering
  \small
  \caption{Candidates with the highest posterior slab-membership probabilities under Sparse RENeGe. Probabilities are approximate because available summaries differ in their final digits. Candidate indices are specific to each network.}
  \label{tab:selected-modes}
  \begin{tabular}{@{}lccc@{}}
    \toprule
    City & First & Second & Third \\
    \midrule
    Barcelona &
    Mode 3 ($\approx0.94$) &
    Mode 16 ($\approx0.94$) &
    Mode 2 ($\approx0.80$) \\
    \addlinespace
    Central Bogot\'a &
    Mode 7 ($\approx0.97$) &
    Mode 2 ($\approx0.91$) &
    Mode 15 ($\approx0.55$) \\
    \bottomrule
  \end{tabular}
\end{table}

\begin{figure}[htbp]
  \centering
  \includegraphics[width=\textwidth]{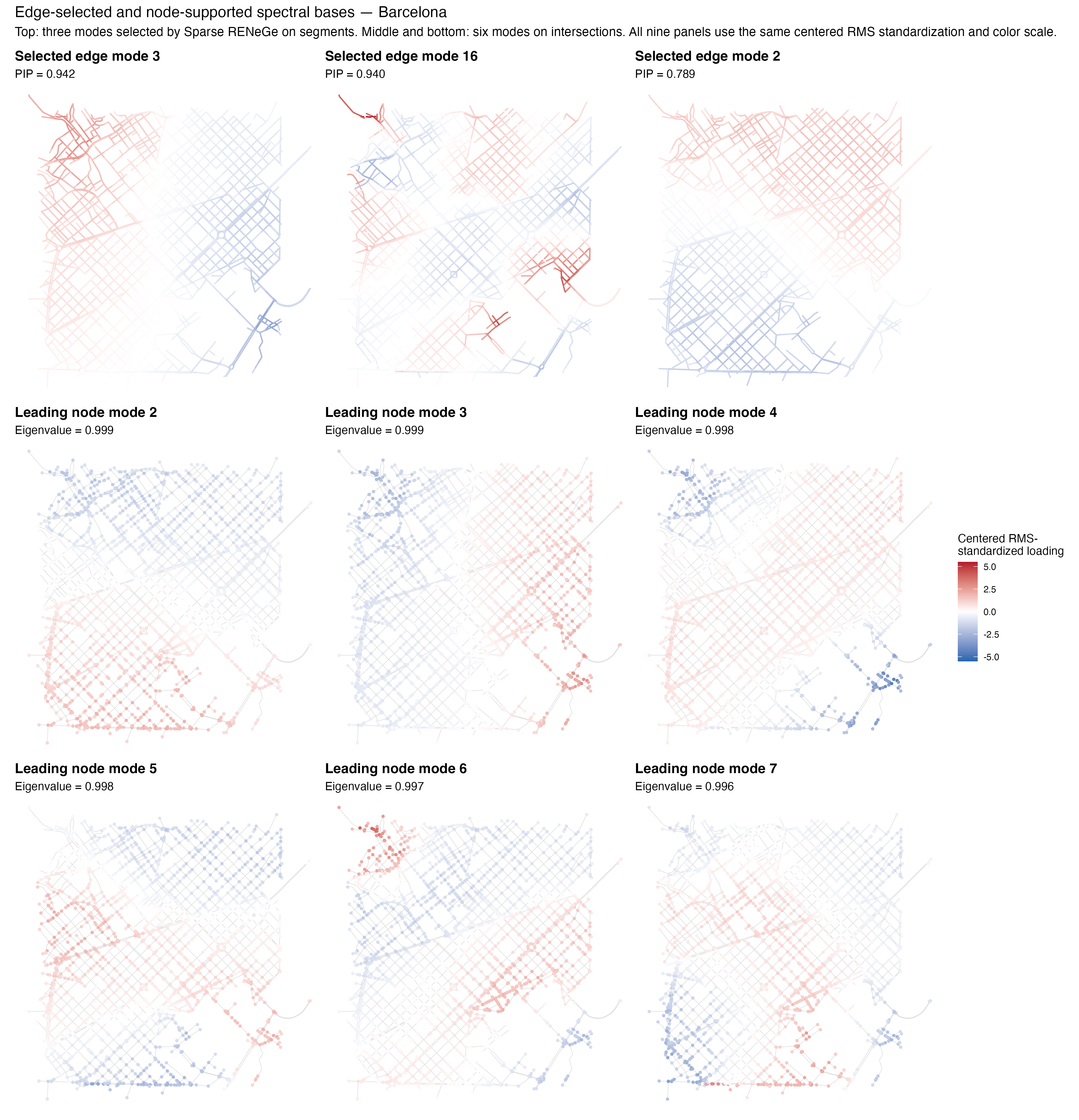}
  \caption{Barcelona edge and nodal spectral basis functions. The upper row shows the three edge candidates with the highest posterior slab-membership probabilities; the remaining rows show six leading nodal functions after Perron-mode exclusion. All columns are centered and RMS-standardized on their respective supports and displayed on a common color scale.}
  \label{fig:edge-node-bases-barcelona}
\end{figure}

\begin{figure}[htbp]
  \centering
  \includegraphics[width=\textwidth]{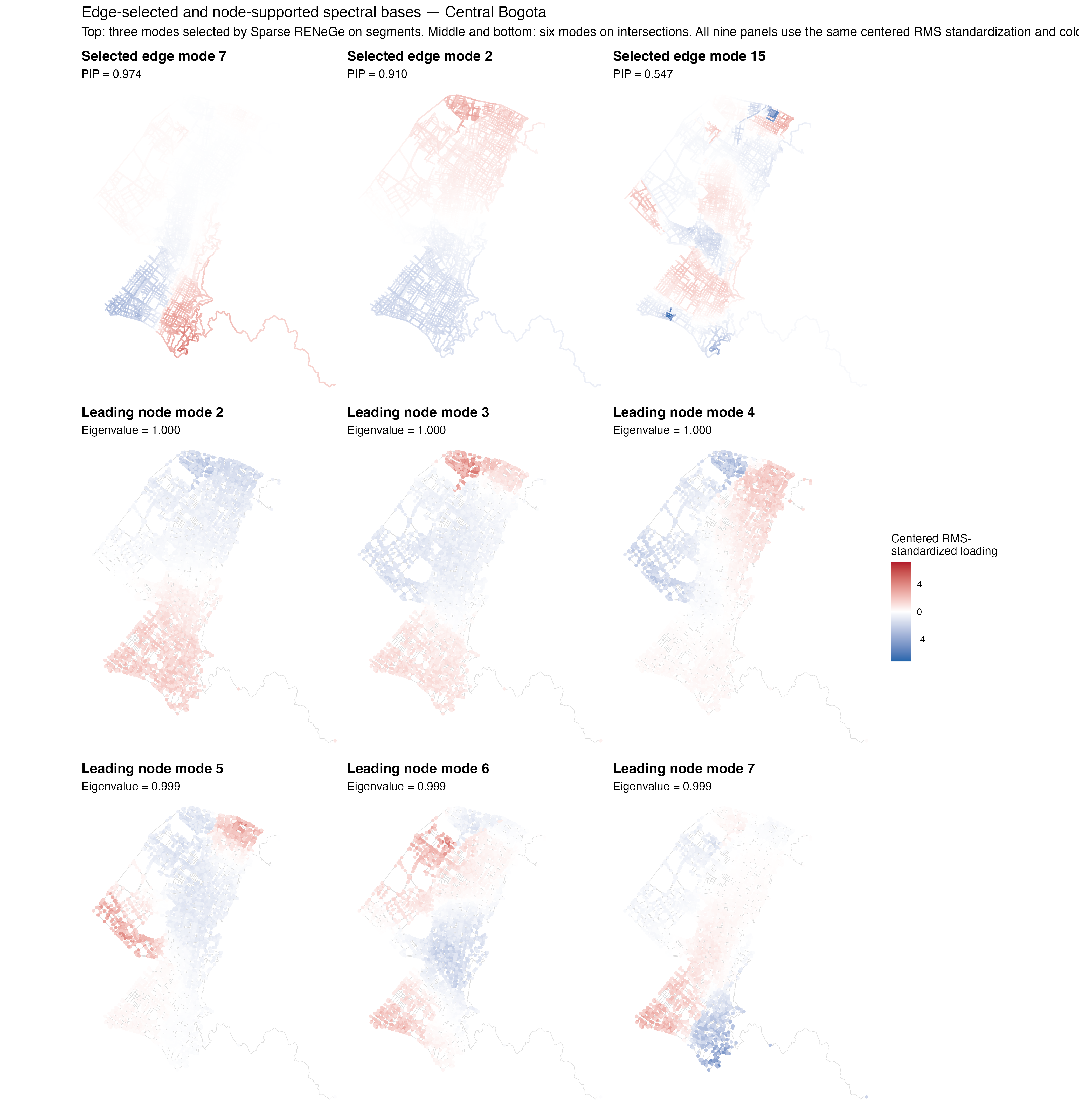}
  \caption{Central Bogot\'a edge and nodal spectral basis functions, arranged and standardized as in Fig.~\ref{fig:edge-node-bases-barcelona}. Edge functions assign values to segments, whereas nodal functions assign values to intersections.}
  \label{fig:edge-node-bases-bogota}
\end{figure}

The coefficient-weighted contributions of these candidates appear in Figs.~\ref{fig:barcelona-contributions} and~\ref{fig:bogota-contributions}. Here, Figs.~\ref{fig:edge-node-bases-barcelona} and~\ref{fig:edge-node-bases-bogota} instead compare their basis functions with six leading nodal spectral functions obtained after excluding the normalized nodal operator's Perron mode. As with the edge operator, the excluded Perron vector need not be constant before degree adjustment. Every displayed column is centered and RMS-standardized on its own support, and the nine panels within each city share a symmetric color scale. The upper row shows edge candidates ranked using crash data; the remaining rows show nodal functions obtained from network geometry without response-based selection.

These displays illustrate how observational support and basis construction affect the available spatial patterns. Edge-supported functions can distinguish incident segments directly, whereas a nodal representation requires a mapping to segment-level effects. Common RMS scaling facilitates visual comparison but does not make the operators, spectral frequencies, or smoothing properties equivalent. Centering also need not preserve eigenvector properties, and the signs of individual basis functions remain arbitrary. Because response-ranked edge candidates are compared with unselected nodal functions, and no nodal model was fitted, the figures do not establish an inferential or predictive advantage for edge-supported models.

\subsection{Derived projections to intersections}
\label{app:intersection-projections}

Intersection-level displays are obtained by applying Eq.~\eqref{eq:posterior-edge-node-map} to the fitted structured segment effects. The projected effect at intersection $v$ is $r_v=d_v^{-1}\sum_{e:\,v\in e}s_e$, where $d_v>0$ is the number of modeled incident segments. Consequently, the displayed posterior mean is
\begin{equation}
  \textsf{E}\big(r_v\mid\boldsymbol{y}\big)
  =
  \frac{1}{d_v}
  \sum_{e:\,v\in e}
  \textsf{E}\big(s_e\mid\boldsymbol{y}\big).
  \label{eq:edge-to-node-projection}
\end{equation}
Here, the projected field is $\widetilde{\mathbf{U}}\boldsymbol{b}$ for Sparse RENeGe, $\sigma_s\boldsymbol{g}$ for spectral CAR, and $\sigma\sqrt{\rho}\,\boldsymbol{g}$ for spectral BYM2. The BYM2 independent component is omitted so that the displays concern network-structured variation in all three models.

\begin{figure}[htbp]
  \centering
  \includegraphics[width=\textwidth]{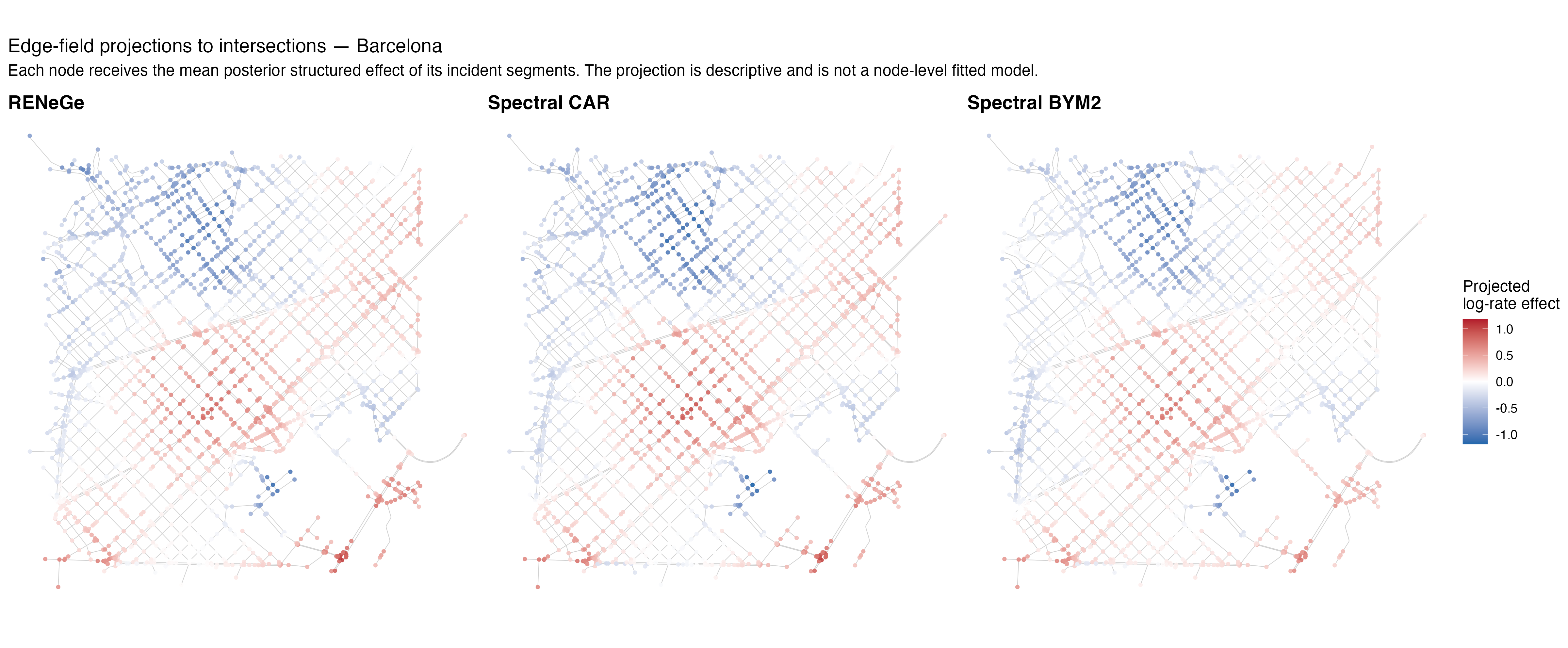}
  \caption{Posterior mean structured segment effects projected to Barcelona intersections using Eq.~\eqref{eq:edge-to-node-projection}. The three panels share a color scale. The BYM2 projection includes only its structured component.}
  \label{fig:node-projection-barcelona}
\end{figure}

\begin{figure}[htbp]
  \centering
  \includegraphics[width=\textwidth]{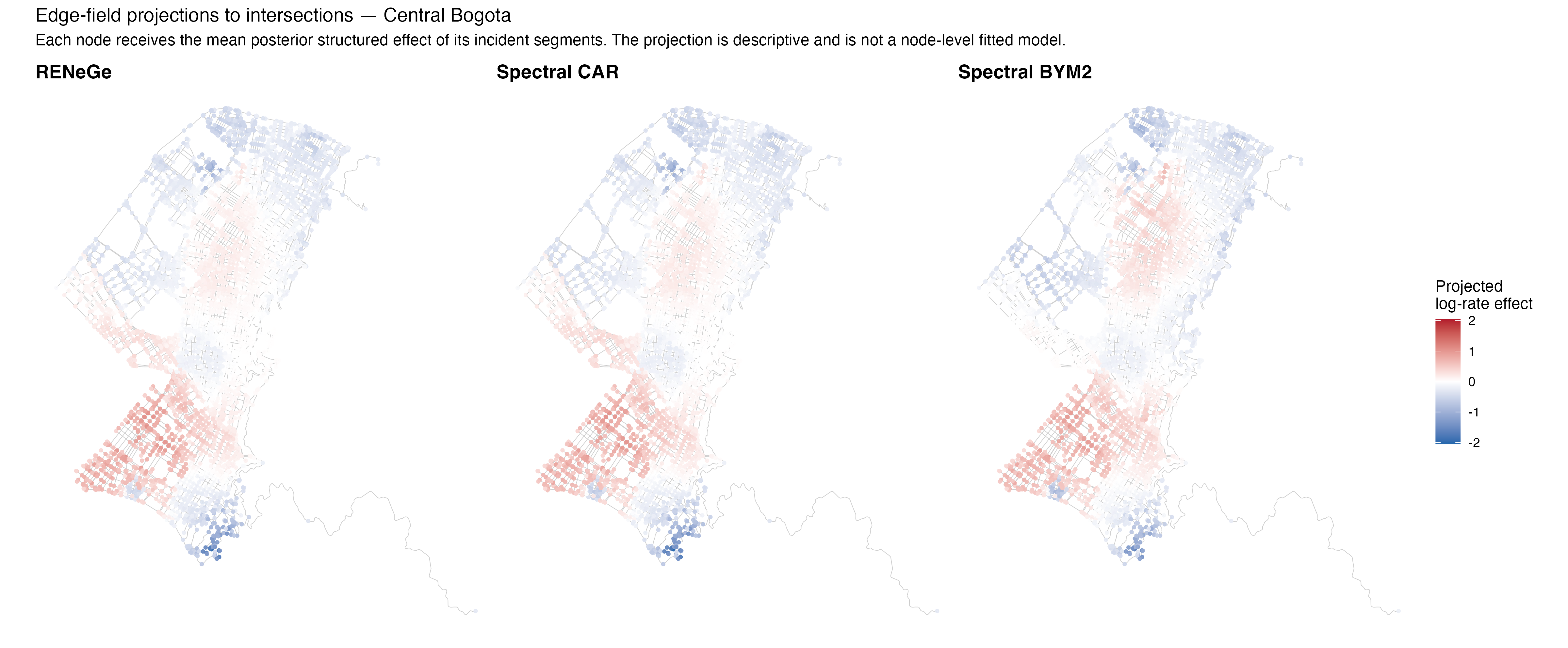}
  \caption{Posterior mean structured segment effects projected to central Bogot\'a intersections. The panels share a color scale and exclude the BYM2 independent component. These are descriptive projections of segment-level fits, not fitted intersection-level models.}
  \label{fig:node-projection-bogota}
\end{figure}

Figures~\ref{fig:node-projection-barcelona} and~\ref{fig:node-projection-bogota} show these projected posterior means on a common color scale within each city. Incident-segment averaging can attenuate local contrasts, and the resulting values remain summaries of latent log-frequency effects rather than intersection-level crash counts or frequencies. Applying the projection to each posterior draw would propagate uncertainty while preserving posterior dependence among incident effects; the displayed mean maps alone do not convey that uncertainty.

\section{Representative Simulated Spatial Fields}
\label{app:simulation-maps}

\begin{figure}[htbp]
  \centering
  \includegraphics[width=\textwidth]{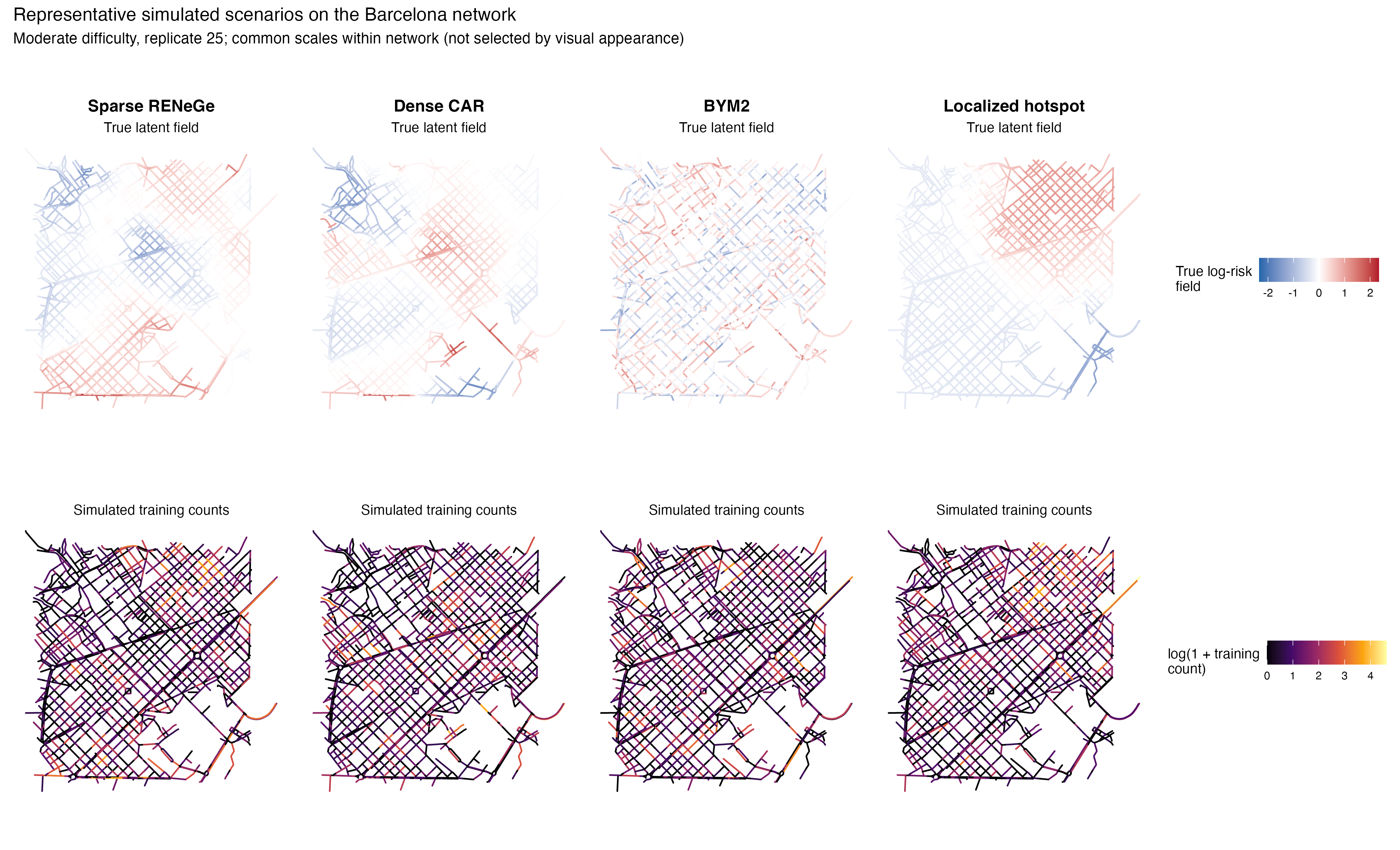}
  \caption{Illustrative moderate-difficulty simulation on the Barcelona network using preassigned replication 25. The upper row shows the true latent log-frequency field under each mechanism; the lower row shows the corresponding training counts on the $\textsf{log}(1+y)$ scale. Color scales are shared within each row.}
  \label{fig:simulation-map-barcelona}
\end{figure}

\begin{figure}[htbp]
  \centering
  \includegraphics[width=\textwidth]{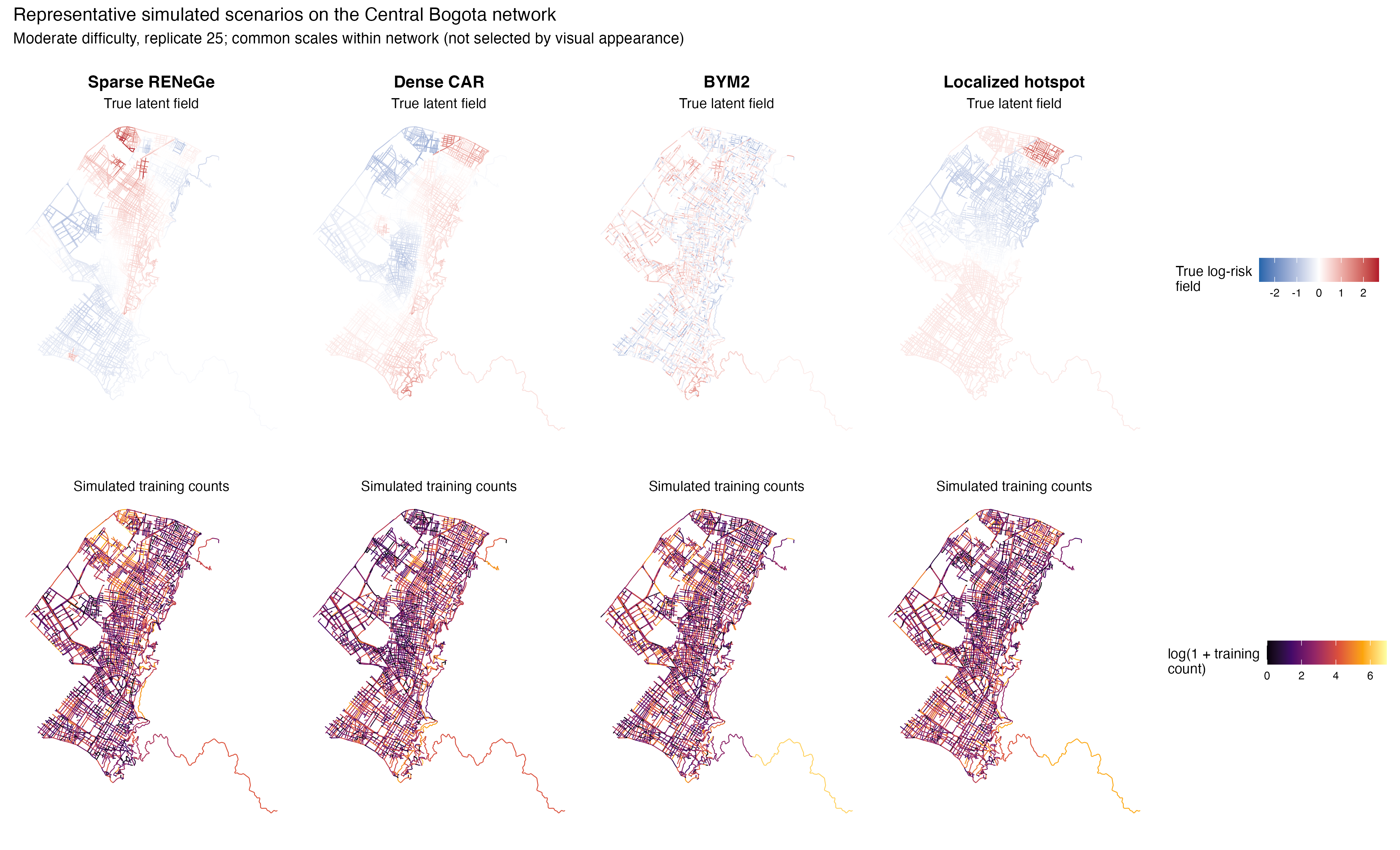}
  \caption{Illustrative moderate-difficulty simulation on the central Bogot\'a network using preassigned replication 25. The upper row shows the true latent log-frequency fields, and the lower row shows the corresponding training counts on the $\textsf{log}(1+y)$ scale. Color scales are shared within each row.}
  \label{fig:simulation-map-bogota}
\end{figure}

Figures~\ref{fig:simulation-map-barcelona} and~\ref{fig:simulation-map-bogota} illustrate the four generating mechanisms on the empirical network geometries. All panels use the preassigned replication 25 under moderate difficulty, avoiding selection based on visual appearance. Within each network, the upper panels share a color scale for the true latent fields, and the lower panels share a separate scale for the corresponding training counts transformed as $\textsf{log}(1+y)$. These examples illustrate the generators and are not intended to summarize variability across replications.

The sparse and dense CAR-type fields show patterns derived from the retained spectral basis, while the BYM2-type field includes additional independent segment-level variation. The localized generator creates positive and negative concentrations in spectral coordinates, which need not correspond to geographically compact hotspots. Differences between the latent-field and count maps reflect both heterogeneous segment exposure and negative-binomial sampling variation. Interpretation of the complete experiment therefore rests on the replicated performance summaries in Section~\ref{sec:simulation-results}, rather than on these individual realizations.

\end{document}